\documentclass[aps,reprint,groupedaddress,notitlepage,amsmath,amssymb,amsfonts,bm,subfigure]{revtex4-2}

\usepackage[utf8]{inputenc}
\usepackage[T1]{fontenc}

\usepackage[american]{babel}
\usepackage[autostyle=true]{csquotes}
\usepackage{graphicx,bm}

\usepackage{xcolor}
\usepackage[normalem]{ulem}

\usepackage[colorlinks=true]{hyperref}
\usepackage[sort&compress, capitalize]{cleveref}

\usepackage[normalem]{ulem}

\newcommand{\norm}[1]{\left\lVert#1\right\rVert}

\begin{document}

\title{Edge manifold as a Lagrangian Coherent Structure in a high-dimensional state space}

\author{Miguel Beneitez$^1$}
\email{beneitez@kth.se}

\author{Yohann Duguet$^2$}
\email{duguet@limsi.fr}

\author{Philipp Schlatter$^1$}
\email{pschlatt@mech.kth.se}

\author{Dan S. Henningson$^1$}
\email{henning@kth.se}
\affiliation{$^1$Linné FLOW Centre and Swedish e-Science Research Centre (SeRC), KTH Engineering Mechanics, Royal Institute of Technology, SE-10044 Stockholm, Sweden\\
$^2$LIMSI-CNRS, Univ. Paris-Saclay, P91405 Orsay, France}
\date{\today}

\begin{abstract}

Dissipative dynamical systems characterised by two basins of attraction are found in many physical systems, notably in hydrodynamics where laminar and turbulent regimes can coexist. The state space of such systems is structured around a dividing manifold called the edge, which separates trajectories attracted by the laminar state from those reaching the turbulent state.  We apply here concepts and tools from Lagrangian data analysis to investigate this edge manifold. This approach is carried out in the state space of automous arbitrarily high-dimensional dissipative systems, in which the edge manifold is re-interpreted as a Lagrangian Coherent Structure (LCS). Two different diagnostics, finite-time Lyapunov exponents and Lagrangian Descriptors, are used and compared with respect to their ability to identify the edge and to their scalability. Their properties are illustrated on several low-order models of subcritical transition of increasing dimension and complexity, as well on well-resolved simulations of the Navier-Stokes equations in the case of plane Couette flow. They allow for a mapping of the global structure of both the state space and the edge manifold based on quantitative information. Both diagnostics can also be used to generate efficient bisection algorithms to approach asymptotic edge states, which outperform classical edge tracking.
\end{abstract}

\maketitle
\section{Introduction}

Many deterministic physical systems can operate in two different regimes depending on initial conditions. This property calls naturally for a geometrical partition of the associated state space in terms of basins of attractions \cite{nusse1996basins}. The identification of the boundaries between basins is crucial for the cartography of the state space, for prediction, as well as for control. Besides such basin boundaries, being invariant sets of the system, also support their own specific dynamics.
Some physical situations where the precise dynamics on the basin boundaries matter, include climate dynamics \cite{lucarini2017edge}, endothermic chemical reactions barriers \cite{junginger2016lagrangian}, synchronisation of phase oscillators \cite{sieber2014controlling}, the instability of accretion disks \cite{rincon2007self},  laser dynamics in modulated optical cavities \cite{marquardt2006dynamical}, magnetic reconnection \cite{cassak2007onset}, free fall  of objects in a gravity field \cite{nave2019global}, chaotic plasma devices \cite{munoz2012}, drift-wave turbulence \cite{Chian2013} and many others. However the main illustration for the study of such a mixed state space comes from hydrodynamics,
more particularly the century-old problem of transition from laminar to turbulence \cite{reynolds1883}.

In the field of hydrodynamics, most incompressible viscous fluid flows near solid boundaries, can undergo transition to turbulence in a subcritical manner \cite{schmid2001stability}. The flow configurations of interest include the flow inside circular or rectangular pipes \cite{eckhardt2007turbulence}, between two plates in motion \cite{daviaud1992subcritical}, inside a rotor-stator \cite{cros2002spatiotemporal} or the flow developing on a semi-infinite flat plate \cite{Kreilos2016}. The laminar state in the examples above is time-independent, characterised by spatial symmetries, and free of fluctuations, while the turbulent state is, depending on the level of modelling, arbitrarily complicated. For the parameters of interest here, the laminar state is linearly stable and coexists with the turbulent set. Some specific trajectories neither reach the laminar state nor the turbulent regime, and have been labelled {\it edge trajectories} \cite{schneider2006edge}. Based on computational attempts to identify edge trajectories, the set of such trajectories has been conjectured to be a differentiable manifold of codimension one in the state space, called either {\it edge of chaos}, {\it laminar/turbulent boundary}, {\it boundary of turbulence} or simply {\it the edge} \cite{Skufca2006,schneider2008edge,duguet2008transition}. At high enough Reynolds numbers, trajectories in the turbulent state are sustained:  both the turbulent and the laminar state are attractors of the system. As far as the Navier-Stokes equations are concerned, this bistable situation was first approached computationally in \cite{itano2001dynamics,toh_itano_2003} who used a simple bisection method to identify the invariant dynamics on the edge, corresponding to a travelling wave solution (a result revised later). The analysis of this orbit and of its instability yields useful information for the understanding of the turbulent dynamics itself, notably regarding bursting dynamics, interpreted as a homoclinic connection between the travelling wave on the edge and a more complicated turbulent-like regime \cite{gibson_halcrow_cvitanovic_2008}.

\begin{figure*}[tb]
    \centering
    \begin{tabular}{c c}
    \includegraphics[width=0.38\textwidth]{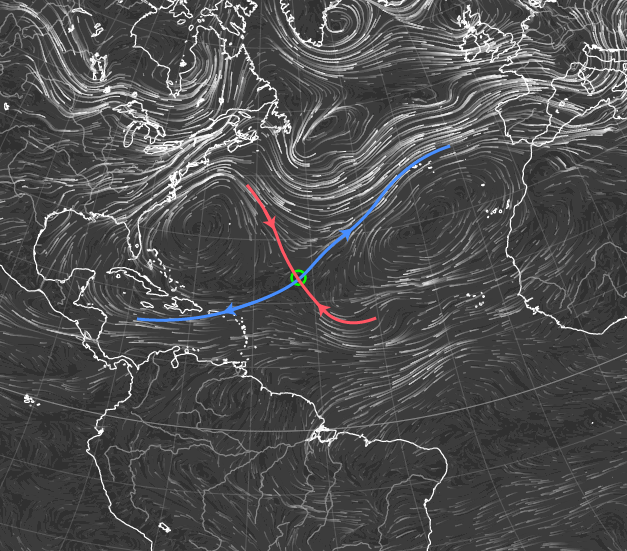}\put(-210,165){$(a)$} &
    \includegraphics[width=0.58\textwidth]{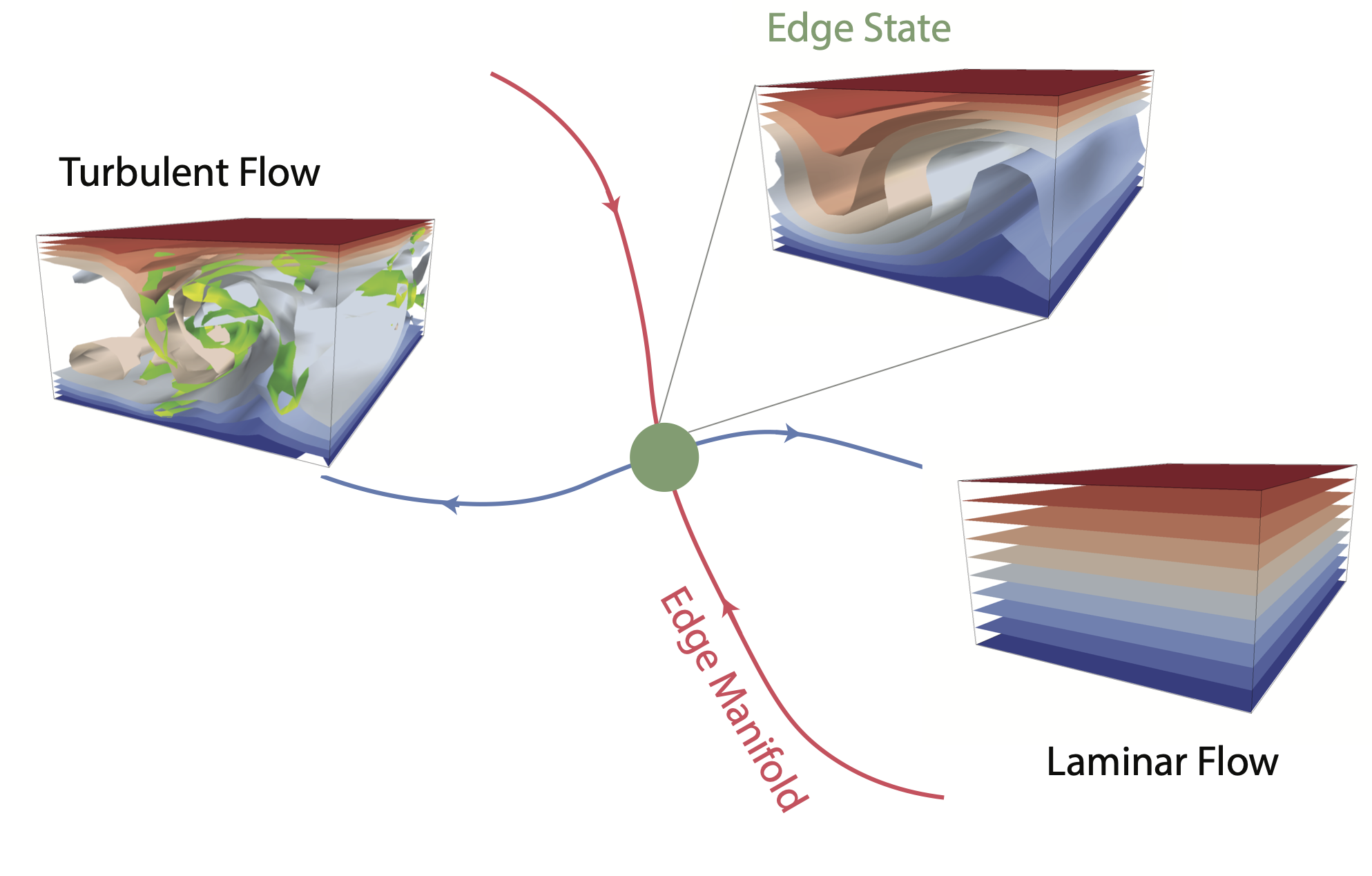}\put(-260,165){$(b)$}
    \end{tabular}
    \caption{Sketch of the analogy between Lagrangian Coherent Structures in conservative systems and the state space geometry of subcritical laminar-turbulent transition in hydrodynamics.  $(a)$: Lagrangian patterns in Nature: wind map on March, 2$^{nd}$ 2020, at a height of 850hPa, adapted from \href{earth.nullschool.net}{earth.nullschool.net}. Sketched attracting/repelling LCS (blue/red resp.) identified with the unstable (resp. stable) manifold of a saddle point (green). $(b)$: Sketch of the state space for a bistable hydrodynamical system, where the stable manifold of the edge state (green) is the basin boundary (red), whereas the unstable manifold (blue) leads to either the laminar or the turbulent attractor. The 3D figures represent flow fields from actual numerical simulations of plane Couette flow. Contours of streamwise velocity in the flow  (red to blue) and an iso-value $\lambda_2$ criterion (green) used for vortex identification.} 
    \label{fig:LCS_illustration}
\end{figure*}

The bistable case is however only an ideal case, because the edge simply coincides with the intersection of the closures of the two basins of attraction. Following \cite{lebovitz2012boundary} we will refer to it as a {\it strong edge}. Ample experimental and numerical evidence has shown that for low enough Reynolds numbers, the turbulent regime is only transient. The notion of basin boundary becomes more fragile and incompatible with the temporally asymptotic notions of stability: the laminar state becomes formally the only attractor of the system. The notion of basin boundary is however routinely generalised to such leaky cases by focusing on the values of the lifetimes associated with each initial condition, at least if the time of residence in the immediate vicinity of turbulent set is
long enough compared to the typical timescales to be specified. We will refer in a such case to a {\it weak edge}. As a result, if the turbulent dynamics is transiently chaotic \cite{moehlis2004} the basin boundary can display a transversely fractal structure, as a consequence of the Lambda-lemma \cite{alligood1996chaos}. edge trajectories remain smooth in all cases and stand out as trajectories with infinite lifetimes. Several algorithms have been constructed in order to approximate such {\it accessible orbits} \cite{grebogi1987basin}, all based on a maximisation of the lifetime \cite{nusse1989procedure,sweet2001stagger,bollt2005path,Skufca2006,schneider2008edge}. In practice the algorithms do not differ much from the standard shooting method. Most importantly, they rely on quantitative criteria globally measured in the state space to distinguish whether a given trajectory has, at a given time, safely entered the basin of attraction of one of the two states. These criteria have so far taken the form of numerical bounds on a given scalar observable. For instance the kinetic energy of the departure from the laminar state is low when the laminar is approached and higher in the turbulent regime itself. Even bounds-based methods, for some flow cases like the Blasius boundary layer flow, have recently shown limitations for the tracking of edge trajectories beyond certain horizon times \cite{beneitez2019edge}. Also, even in the general cases the identification of relevant bounds for a well-chosen observable implies a very good knowledge of the system that is not always available. Moreover, the outcome of a classical bisection process is a list of Booleans: little information about neighbouring trajectories is saved apart from the \textquote{laminar} or \textquote{turbulent} label at large enough times. Critically, the very notion of edge is defined so far both by the algorithmic way to identify the edge and by the infinite-time notions of stability found in usual textbooks. A different paradigm, based on quantitative criteria, optimal quantities and on a finite-time framework, is hence necessary for a more transparent and universal characterisation of edge manifolds.

Beyond the Boolean bounds-based definition of the edge manifold itself, other important properties can be quantified. The long-time outcome of classical bisection itself reveals a third asymptotic regime, unstable by construction but specific to the dynamics within that invariant manifold. This asymptotic regime, the {\it edge state} is the relative attractor on the codimension one manifold. Depending on the flow case and the symmetries of the system (discrete and continuous), an equilibrium, a relative equilibrium (a travelling wave), a periodic orbit, a relative periodic orbit, a torus or even a chaotic attractor can be edge states. In each of these cases, the invariant manifold appears by definition as the set of initial conditions that converge in forward time to the edge state: defining the edge manifold as the stable manifold of the edge state. Complications do arise in some cases, seen in practice, where more than one distinct edge states exist for the same parameter values \cite{duguet2008transition,khapko2014complexity}. The edge can then be globally interpreted as the closure of the union of such stable manifolds.

A few specific parts of the edge manifold are also important: by assuming a given distance defined in the entire state space, the minimal seed is defined loosely as a state on the edge minimising the distance to the laminar state \cite{Duguet2013,Kerswell:2018aa}. The notion of minimal seed (and the associated minimal distance) is useful in assessing the stability of the laminar state with respect to finite-amplitude disturbances.
Recently, it was demonstrated that the edge state and the turbulent state often arise from the same saddle-node bifurcation followed by further bifurcations \cite{Kreilos2012}. Even the leaky property of the turbulent state, when present, emerges in boundary crises involving the collision between the turbulent state and the edge manifold as parameters are varied \cite{avila2013streamwise,zammert2015crisis}. This makes the identification of edge states instrumental for the determination of the full, exact bifurcation diagram of the system.

Interpreting the edge manifold as a stable manifold of some object to be found suggests bridges to other areas of dynamical systems. Many different tools have been developed and analysed recently to identify the locally most repelling (resp. attracting) material surfaces. This forms the concept of Lagrangian Coherent Structure (LCS), where {\it Lagrangian} refers to the tracking of individual trajectories \cite{Haller_2015}. LCSs correspond often in applications to stable (and unstable) manifolds of fixed points, as in Fig.\ \ref{fig:LCS_illustration}(a). The original frame for which they have been developed is also inspired by hydrodynamics: it deals with the finite-time transport of fluid particles by known time-dependent fluid flows \cite{haller2000lagrangian,peacock2013lagrangian}. Other applications, for instance chemical reactions, have also been considered \cite{junginger2016lagrangian}. Mathematically yet, the class of systems under study is radically different: the dynamics are always conservative, there are no attractors, the trajectories can be computed in either forward or backward time, and the dimension of the state space (which coincides with the physical space) is very low, typically two. Besides, the system is in general non-autonomous due to the time dependence of the vector field. In most applications of Lagrangian concepts, the state space associated with Lagrangian tracers (governed by an equation $\dot{x}=v$) and the physical space coincide. In the Eulerian point of view relevant to the transition problem, both spaces differ radically in their dimension.

Despite these differences, the goal of this article is to demonstrate that the toolbox developed over the years for the study of LCSs can be used for the investigation and the numerical identification of the edge manifold in transitional flows according to the analogy expressed in Fig.\ \ref{fig:LCS_illustration}. It is in line with the point of view used first by Ref.\ \cite{aldridge2006} for  multi-dimensional dissipative systems. Dissipative PDEs formally have a state space of infinite dimension (in practice their numerical discretization still yields a finite dimension of $\mathcal{O}(10^5)$ or more). The scalability property of the tools considered is hence crucial for the feasibility of the whole method. However we restrain the analysis to autonomous velocity fields, this being the generic situation in all the hydrodynamic models considered here. We focus in particular mainly on two distinct tools popular in the recent literature on LCS: finite-time Lyapunov exponents (FTLEs) \cite{Haller_2015} and Lagrangian Descriptors (LDs) \cite{mancho2013lagrangian,mendoza2014lagrangian}. We give key examples of application of these tools in subcritical shear flow models of increasing complexity and dimensionality, following a hierarchy which ranges from a two-dimensional state space to infinite dimension in the incompressible Navier-Stokes equations. Using the presented framework we suggest and test the use of these new diagnostic to improve edge tracking algorithms.


\section{Definitions and LCS diagnostics}
\label{definitions}
We consider here a general dynamical system defined on a forward-invariant subset $A\subset \mathbb{R}^n$, governed by
\begin{equation}
    \dot{x} = f(x,t), \quad x\in A, \quad t\in  [t_0,t_0+\tau], \label{eq:vectorField}
\end{equation}
where $f: A\times [t_0,t_0+\tau] \to \mathbb{R}^n$ is a sufficiently smooth vector field and $\tau>0$. Let $F_{t_0}^{t}$ be the flow map
\begin{equation}
\begin{split}
    F_{t_0}^{t}:\ & A \to A\\
    & x_0 \to x(t;t_0,x_0),
\end{split}
\end{equation}
which maps an initial position $x_0$ at time $t_0$ to its position at time $t$. Linearizing \eqref{eq:vectorField} around the trajectory $x(t)\equiv x(t;t_0,x_0)$ reads
\begin{equation}
\dot{z} = \nabla_x f(x(t),t) z, \quad z\in \mathbb{R}^n, \quad t\in [t_0,t_0+\tau],
\end{equation}
where $z$ is the linearized solution and $\nabla_x f(x(t),t)$ is the Jacobian of the vector field $f$. The deformation gradient $\nabla F_{t_0}^t$ corresponds to the fundamental solution matrix of the equation of variations so that
\begin{equation}
    z(t;t_0,z_0) = \nabla F_{t_0}^{t}(x_0)z_0,
\end{equation}
in the interval $t\in [t_0,t_0+\tau]$.
\subsection{Finite-Time Lyapunov exponents}

Finite-time Lyapunov exponents are a popular tool for the identification of LCS (cf. Ref.\  \cite{Haller_2015} and the references in Section 4.2). Let $C_{t_0}^{t}$ be the (positive definite) Cauchy-Green tensor
\begin{equation}
    C_{t_0}^{t} = (\nabla F_{t_0}^{t})^{*}\nabla F_{t_0}^{t},
\end{equation}
where $(\cdot)^*$ represents the transpose. Let $\lambda_i~i=1,...,n$ denote the eigenvalues of $C_{t_0}^{t}$ and $\xi_i$ its associated eigenvectors such that
\begin{equation}
    C_{t_0}^{t}\xi_i = \lambda_i\xi_i,\quad \left|\left|\xi_i\right|\right|=1,\quad i=1,\dots,n,
\end{equation}
and
\begin{equation*}
    \lambda_1\ge \dots \ge \lambda_n \ge 0,\quad \xi_i \perp \xi_j,\quad i\neq j,
\end{equation*}
where $\left|\left| {\cdot} \right|\right|$ denotes the $L^2$-norm. The maximum of all factors by which small perturbations vectors are stretched over the time interval $(t_0,t_0+\tau)$ around an initial condition $x_0$ is given by $\sqrt{\lambda_1(t_0,t_0+\tau,x_0)}$. The $i^{th}$ finite-time Lyapunov exponent of the system at position $x_0$ is given by
\begin{equation}
    \Lambda_i(t_0,t_0+\tau,x_0) = \frac{1}{\tau}\log\sqrt{\lambda_i(t_0,t_0+\tau,x_0)},
\end{equation}
for $i=1,\dots,n$. The  ridges in the field of the largest FTLE at time $t_0+\tau$ (labelled simply $\lambda_{t_0}^{t_0+\tau}$) can be used as a diagnostic of hyperbolic LCS \cite{HALLER2011574,Haller_2015,Hadjighasem_2017}. Ref.\ \cite{HALLER2011574} provides rigorous theorems that establish a precise connection between hyperbolic LCS and FTLE ridges when further conditions on the rate-of-strain tensor are satisfied along the ridges. The Cauchy-Green tensor has size $n\times n$, and its entries are usually computed using second-order centered finite-differences \cite{Haller_2015}. Hyperbolic LCSs refer to attracting and repelling distinguished invariant manifolds \cite{haller2000lagrangian}, with forward time FTLEs relevant for the identification of repelling LCS, and backward time FTLEs relevant for the identification of attracting LCS \cite{haller2011sapsis}.

The computational cost of FTLEs following the above method increases rapidly with the state space dimension and is unavoidably costly for large dimensions. FTLEs can however be computed in arbitrarily high dimension using the recent algorithm based on Optimally Time-Dependent (OTD) modes, which are determined from a minimization principle \cite{babaee2016minimization}. Under generic conditions these OTD modes converge exponentially fast to the eigendirections of the Cauchy-Green tensor associated with the largest eigenvalues, i.e.\ the dominant finite-time Lyapunov exponents \cite{babaee2017}. In practice for the present LCS diagnostic, only the largest exponent needs to be computed, for which other methods are available.

\subsection{Lagrangian Descriptors}

Lagrangian Descriptors (LDs) are a more recent diagnostic for Lagrangian coherence which does not require a differentiation of the flow map with respect to the initial condition. This diagnostic was introduced in \cite{mendoza2010hidden,mancho2013lagrangian} and further developed in \cite{lopesino2017theoretical}.
LDs are based on the integration of a given observable along trajectories. The original quantity of interest is
\begin{equation}
    M(x_0,t_0,\tau)=\int_{t_0-\tau}^{t_0+\tau} g(x(t))dt,\label{eq:Mfull}.
\end{equation}
In Eq.\ \eqref{eq:Mfull}, the observable $g$ is taken as
\begin{equation}
g(x(t))=\sum_{i=1}^{m} |f_i(x,t)|^{p},
\label{eq:g}
\end{equation}
where the $f_i$'s are the components of the velocity field $f$, and $p\in(0,1]$ and $\tau\in \mathbb{R}^{+}$ are two parameters.
We focus on the definition \eqref{eq:Mfull} used in the literature \cite{naik2019finding, naik2019NHIM} although other alternative definitions have been suggested. It is convenient to split Eq.\ \eqref{eq:Mfull} into its forward and backwards contributions  \cite{junginger2016lagrangian,naik2019finding}:
\begin{align}
   M(x_0,t_0,\tau)^+&=\int_{t_0}^{t_0+\tau}g(x(t))dt,\\
   M(x_0,t_0,\tau)^-&=M-M^+.
\end{align}
Due to the dissipative nature of Eq.\ \eqref{eq:vectorField}, numerical backwards integration is ruled out for stability issues and only $M^+$ can be considered here. Since our focus is on stable manifolds rather than unstable ones it is sufficient to focus on the computation of $M^+$. LDs have been used to identify boundaries between qualitatively different dynamics, based on abrupt variations of $M^+$. Alternatively, it is useful to quantify the abrupt changes of $M^+$ and thus to consider the Euclidian norm of its gradient
\begin{equation}
    B(x_0,t_0,\tau) = \left[\sum_{i=1}^n\left(\frac{\partial M^+}{\partial x_{0,i}}(x_0,t_0,\tau)\right)^2\right]^{1/2}, \label{eq:Bfull}
\end{equation}

Many other diagnostics for LCS have been suggested, see \cite{Hadjighasem_2017} for a recent comparative review. Alongside the diagnostics based on a scalar field, such as FTLEs and LDs, other approaches based on transfer operators or dynamic Laplace operators seek coherent structures by formulating rigorous mathematical coherence principles (see e.g.\ \cite{allshouse2015lagrangian,Hadjighasem_2017}). These diagnostics generally display limited scalability properties and are not considered here.

\section{LCS identification of the edge}
\label{modelsSection}
In this section we demonstrate that the edge is highlighted as an LCS for several nonlinear models of increasing complexity, from two-dimensional models to thousands of degrees of freedom in the Navier-Stokes equations. We show that the edge can be effectively identified in state space even when its identification using global observables can be challenging.

\subsection{Hierarchy of low-order shear flow models} \label{models}

The hierarchy of low-order models on which the LCS indicators are tested is based on a Galerkin truncation of the Navier-Stokes equations, in the spirit of the derivation of the Lorenz model \cite{lorenz1963deterministic}. We assume that the bold vector ${\bm X}$ $\in \mathbb{R}^3$ represents the position in the physical space coordinates (which differ from the state space coordinates), that the velocity field ${\bm v_{b}}$ is the stable base flow solution, i.e.\ a steady solution of the governing PDEs. Let ${\bm u}={\bm v}-{\bm v_b}$ be the perturbation velocity to the base flow, not to be confuse with the tangent field $f$. At any time, each model assumes that ${\bm u}$ can be written as
\begin{equation}
    {\bm u}({\bm X},t)=\sum_{i=1}^{n}a_i(t)\hat{\bm u}_i({\bm X}),
    \label{Galerkin}
\end{equation}
with $n$ the state space dimension and $\hat{\bm  u}_i,~i=1,..,n$ a basis of predetermined vector fields. The vector $x(t) \in \mathbb{R}^n$ contains all the amplitudes $a_i,~i=1,...,n$. We consider that for all models, Eq.\ \eqref{eq:vectorField} can be written under the generic form
\begin{equation}
    \dot{x}= Lx+N(x), \label{eq:modelsGeneral}
\end{equation}
with $x\in \mathbb{R}^n$, $L$ an $n\times n$ linear operator and $N$ a quadratic form containing the nonlinear terms. As explained in \cite{waleffe1997self}, all models of subcritical transition consistent with the original PDEs are subject to two constraints: i) $L$ is a non-normal operator ($LL^* \neq L^*L$) with a stable eigenspectrum and ii) the nonlinear terms do not contribute to the change in energy, i.e.\ $\langle N(x),x \rangle=0$ for all $x$, where $\langle\cdot,\cdot \rangle$ defines an inner product.

\subsection{Dauchot-Manneville model (DM2D)} \label{DM2D}

\begin{figure}[tb]
    \centering
    \includegraphics[width=\columnwidth]{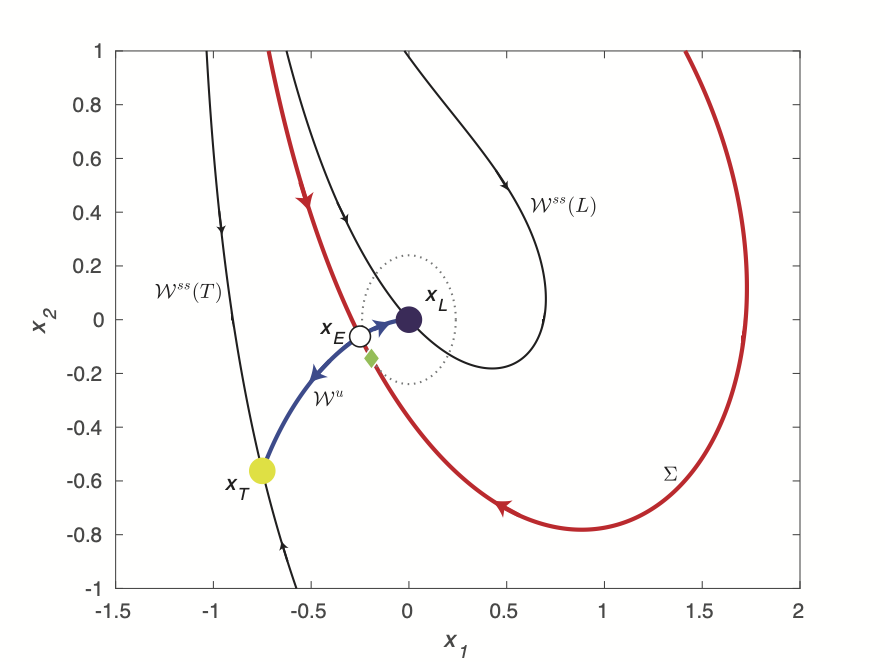}
    \caption{State portrait of the DM2D model for $s_1=-0.1875$ and $s_2=-1$. Filled circles: attracting fixed points, open circle: edge state. The green diamond on $\Sigma$ denotes the minimal seed.}
    \label{fig:DM2D_stateSpace}
\end{figure}

\begin{figure*}[tb]
    \centering
    \begin{tabular}{cc}
        {\includegraphics[width=\columnwidth]{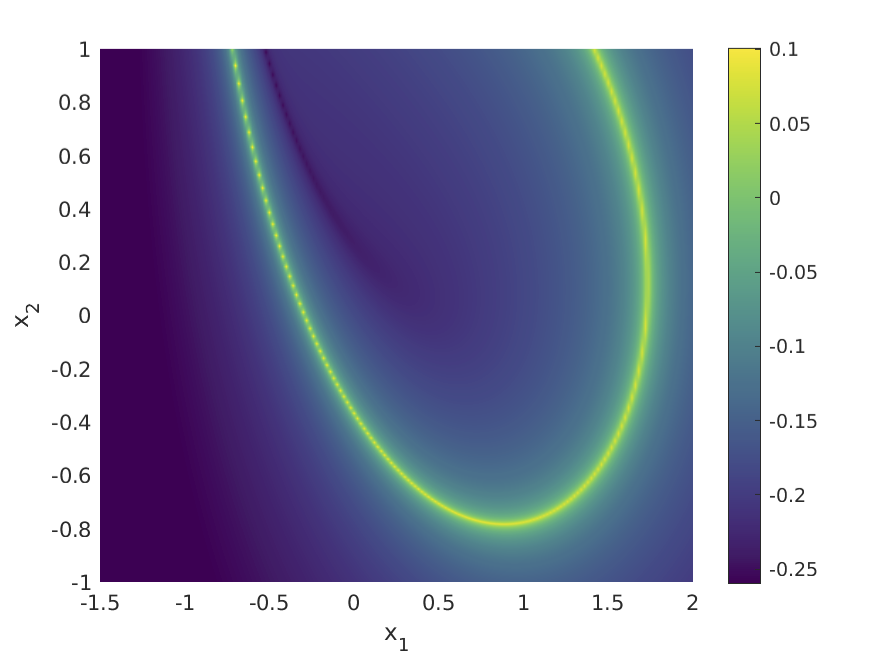}}\put(-230,180){$(a)$} &
        {\includegraphics[width=\columnwidth]{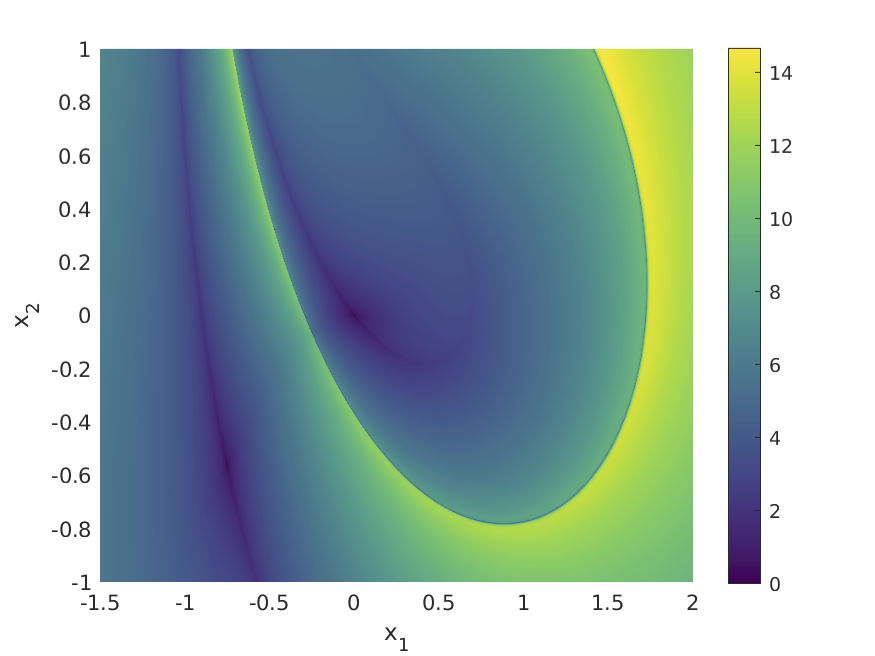}}\put(-230,180){$(b)$}\\
        {\includegraphics[width=\columnwidth]{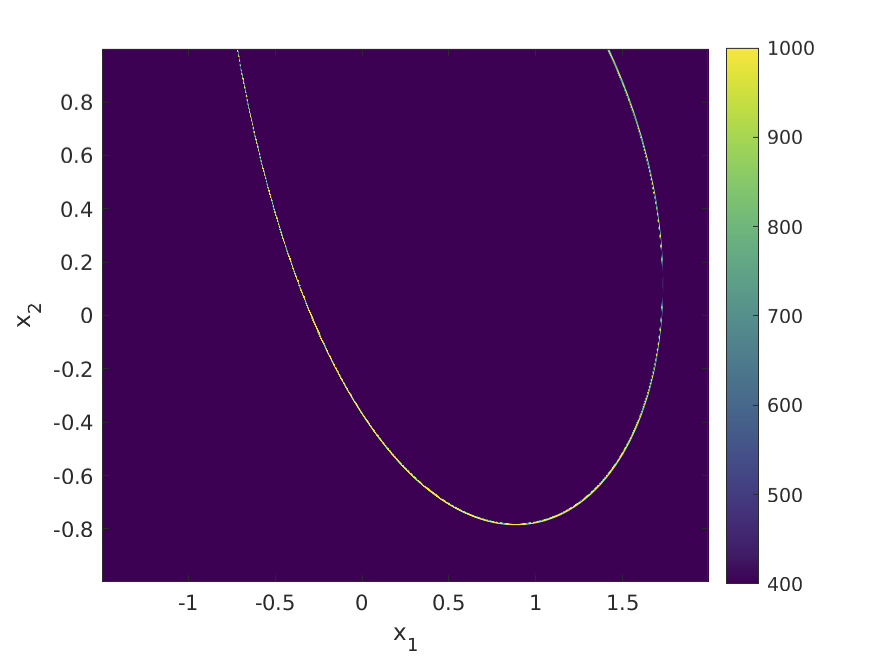}}\put(-230,180){$(c)$} &
        {\includegraphics[width=0.95\columnwidth]{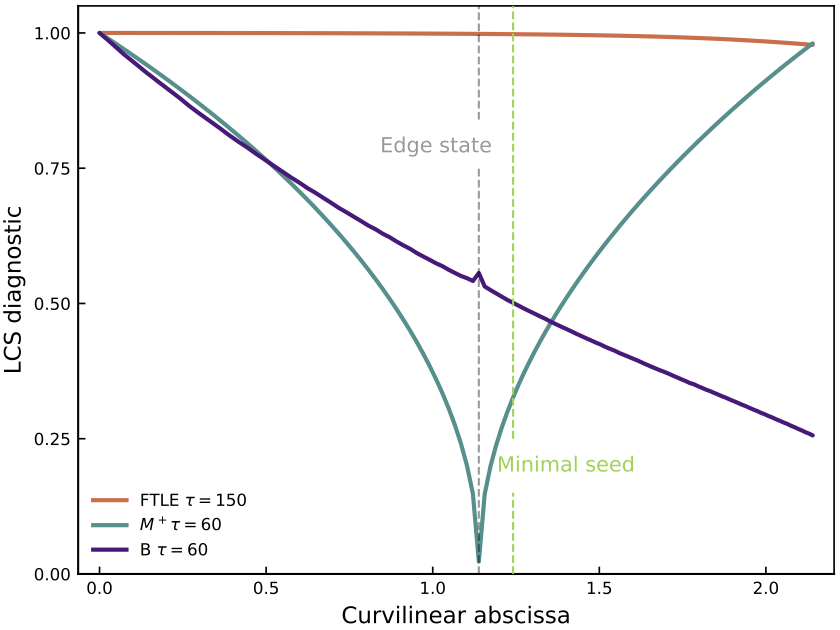}}\put(-230,180){$(d)$}
    \end{tabular}
    \caption{LCS diagnostics for DM2D. $(a)$ largest Finite-time Lyapunov exponent (FTLE) for $\tau=60$. $(b)$ Forward-time Lagrangian Descriptor $M^+$ for $p=0.5$ and $\tau=60$. $(c)$ $B$ for $p=0.5$ and $\tau=60$. $(d)$ LCS diagnostics along part of the ridge in (a)-(c) starting at $(-0.71,0.9733)$ and ending at $(0.5,-0.7051)$. All curves are normalised to $1$ at their starting point.}
    \label{fig:DM2D_diagnostics}
\end{figure*}

As an illustrative case, we consider the simple two-dimensional model introduced by Dauchot and Manneville \cite{Dauchot_1997}, henceforth referred to as DM2D. Although introduced originally in the context of hydrodynamic stability, its typical phase portrait (see Fig.\ \ref{fig:DM2D_stateSpace}) appears in many different applications, e.g.\ in chemistry for potential barriers \cite{junginger2016lagrangian}, in ecology for the competition of two species \cite{strogatz2001nonlinear} or in mechanics for the free fall of objects \cite{nave2019global}. There is no chaos in this model, only fixed points as attractors and one saddle point as edge state. The matrix $L$ and the vector $N$ terms read respectively
\[
L=\left[\begin{array}{cc}
    s_1 & 1 \\
    0   & s_2
\end{array}\right],\quad
N(x_1,x_2)=\left[\begin{array}{c}
     x_1x_2  \\
     -x_1^2
\end{array}\right].
\]
$s_1$ and $s_2$ are two negative parameters, so that the laminar state $x_L=(x_1^L,x_2^L)=(0,0)$ is linearly stable. The main control parameter is the discriminant $\Delta=1-4s_1s_2$, for which $x_L$ is the only fixed point if $\Delta <0$. Two additional fixed points $x_E$ and $x_T$ appear in a saddle-node bifurcation for $\Delta \ge 0$, given by
\begin{eqnarray}
    x_E &=\left( \frac{1}{2}(-1 + \sqrt{\Delta}),\frac{1}{4s_2}(-1 + \sqrt{\Delta})^2\right), \\
    x_T &=\left( \frac{1}{2}(-1 - \sqrt{\Delta}),\frac{1}{4s_2}(-1 - \sqrt{\Delta})^2 \right).
    \label{eq:xDM2D}
\end{eqnarray}
For $0\leq\Delta<1$ the model is bistable: it features two well-defined basins of attraction. They are separated by a smooth edge manifold $\Sigma=\mathcal{W}^s(\{x_E\})$ of the strong type.  The saddle point $x_E$ is the edge state, whereas $x_L$ and $x_T$ are attractors interpreted as the laminar and turbulent state, respectively. Fig.\ \ref{fig:DM2D_stateSpace} shows a phase portrait of the model for the parameters $s_1=-0.1875$ and $s_2=-1$.
Two LCS diagnostics, FTLEs and LDs, are applied to this model with a time horizon fixed to $\tau =60$. For the FTLE field, a clear ridge is visible in Fig.\ \ref{fig:DM2D_diagnostics}(a). Fig.\ \ref{fig:DM2D_diagnostics}(d) shows the largest FTLE along part of $\Sigma$ -- normalised to $1$ at the curvilinear abscissa = 0 --, indicating a smooth and uniform variation of the FTLE along the ridge.
As for the LDs, the contours of $M^+$ in Fig.\ \ref{fig:DM2D_diagnostics}(b) highlight a state space structure perfectly conform to the phase portrait of Fig.\ \ref{fig:DM2D_stateSpace}. There is a non-differentiable minimum of $M^+$ across $\Sigma$, as in former applications of LDs to non-dissipative systems. Isocontours of its gradient norm $B$ happens to be convenient for plotting purposes (\ref{fig:DM2D_diagnostics}(c)). $B$ is computed using centered finite differences:
\begin{multline}
    B = \left[\sum_{i=1}^2\left(\frac{\partial M^+(x_0,t_0,\tau)}{\partial x_{0,i}}\right)^2\right]^{1/2} \approx \\ \left[\sum_{i=1}^2\left(\frac{M^+(x_0+\varepsilon_i),t_0,\tau)-M^+(x_0-\varepsilon_i,t_0,\tau)}{2\varepsilon_i}\right)^2\right]^{1/2},\label{eq:discreteB}
\end{multline}
where $\varepsilon_i$ is a small variation along each state space direction. $B$ is shown in Fig.\ \ref{fig:DM2D_diagnostics}(c). The isolevels of $B$ highlight the dip of $M^+$ map in Fig.\ \ref{fig:DM2D_diagnostics}(b) as a singular feature that coincides with $\Sigma$. The variations of $M^+$ and $B$ \emph{along} the edge are quantified in Fig.\ \ref{fig:DM2D_diagnostics}(d). There is a non-smoooth minimum (actually a zero) of $M^+=0$  coinciding with the saddle fixed point $x_E$. The non-smooth minimum shows a discontinuity at the same location for $B$. The minimal seed does not show any distinctive feature along $\Sigma$. The influence of $p \le 1$ has been well studied in \cite{lopesino2017theoretical}. It is mainly concerned with type of singularity. $p=1$ favours a linear behaviour for $M^+$, however the ridge in $B$ remains.  A 1D cut along the LD maps is shown in Fig.\ \ref{fig:LD_landscapeComp}(a) and Fig.\ \ref{fig:DM2D_bisectionComparison}(b) for the discussion in Section \ref{edgetracking_difmeth}. The ridge in the FTLE (or LD) maps becomes thinner with increasing horizon time $\tau$.
From each of these two diagnostics alone, the edge manifold is hence identified as a repelling LCS once the horizon time is long enough.

\subsection{Lebovitz-Mariotti model (LM6D)}
\label{LM6D}
Increasing the complexity of the system to six dimensions and non-trivial attractors, we consider the model introduced in the work of Lebovitz and Mariotti \cite{lebovitz2013edge}, from this point called LM6D. The model was originally suggested to illustrate concepts of complex boundaries and boundary collapse inspired by transient chaos in hydrodynamics \cite{hof2006finite,eckhardt2007turbulence}.
The derivation of the model is similar to that of \cite{waleffe1997self} and the details of the derivation are introduced in \cite{mariotti2011low}. The matrix $L$ is such that $L_{ij}=-k_i\delta_{ij} + \sigma_0\delta_{i2}\delta_{j3} + \sigma_3(\delta_{i6}\delta_{j4}-\delta_{i4}\delta_{j6})$ and the vector $N$ reads
\[
N(x)=\left[\begin{array}{c}
    -\sigma_0x_2 x_3  \\
    \sigma_0 x_1 x_3-\sigma_1 x_4 x_5 \\
    -(\sigma_4+\sigma_5)x_5 x_6 \\
    \sigma_2 x_2 x_5-\sigma_3 x_1 x_6 \\
    (\sigma_1-\sigma_2)x_2 x_4+(\sigma_4-\sigma_6) x_3 x_6 \\
    (\sigma_5+\sigma_6)x_3 x_5+\sigma_3 x_1 x_4
\end{array}\right],
\]
where $k_i$, $i=1,\dots,6$ and $\sigma_i$ $i=0,\dots,6$ are parameters of the model \footnote{The parameters $k_i$, $i=1,\dots,6$ and $\sigma_i$ $i=0,\dots,6$ are functions of the wavenumbers $\alpha$ and $\gamma$, determined from the formulas in \cite{mariotti2011low}. The values of taken here are $\alpha=1.1,\beta=\pi/2,\gamma=5/3$. The parameter $R$ is akin to the Reynolds number in fluid mechanics and chosen to be $R=315$}.
\begin{figure*}[tb]
    \centering
    \begin{tabular}{cc}
        {\includegraphics[width=\columnwidth]{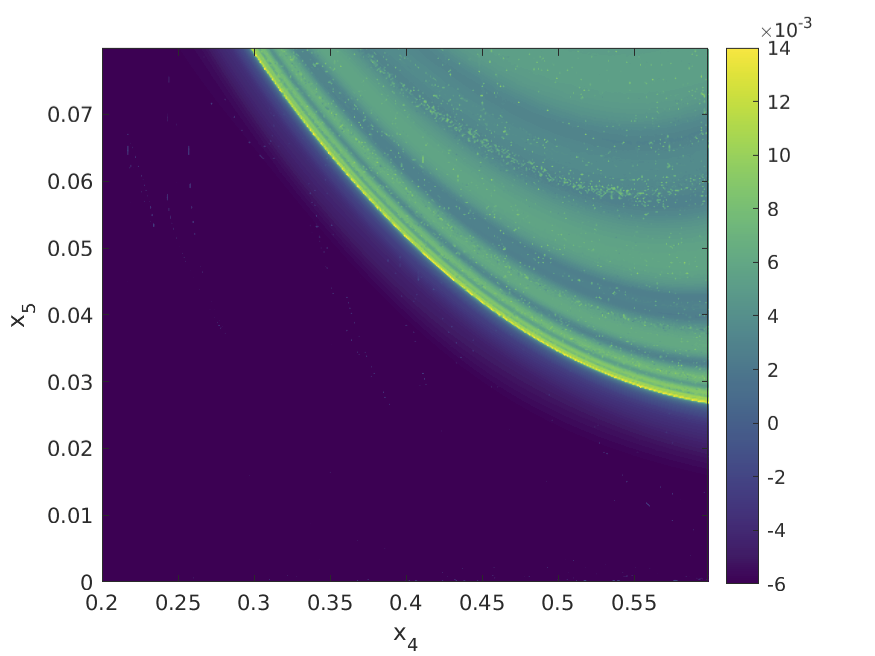}}\put(-230,180){$(a)$} &
        {\includegraphics[width=\columnwidth]{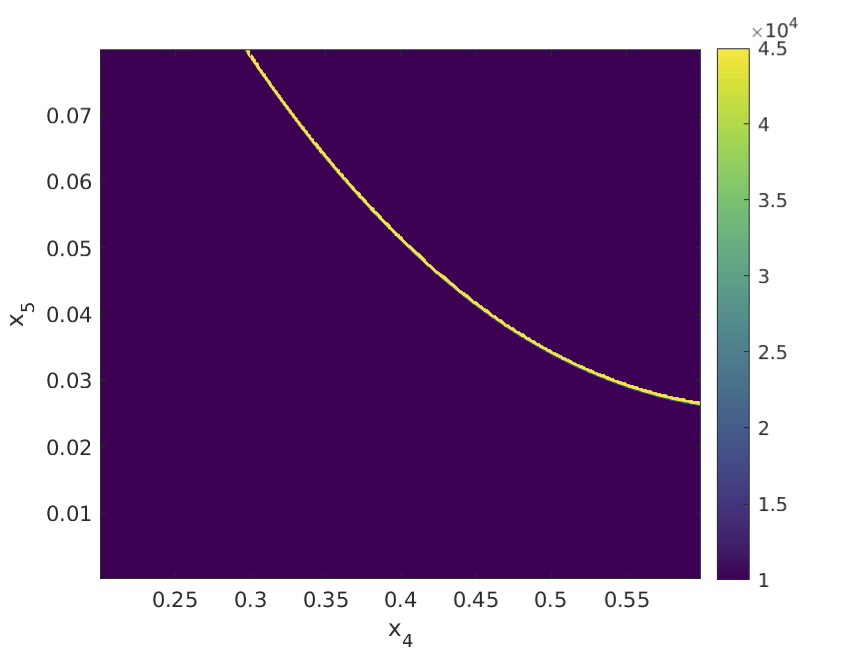}}\put(-230,180){$(b)$}\\
        {\includegraphics[width=\columnwidth]{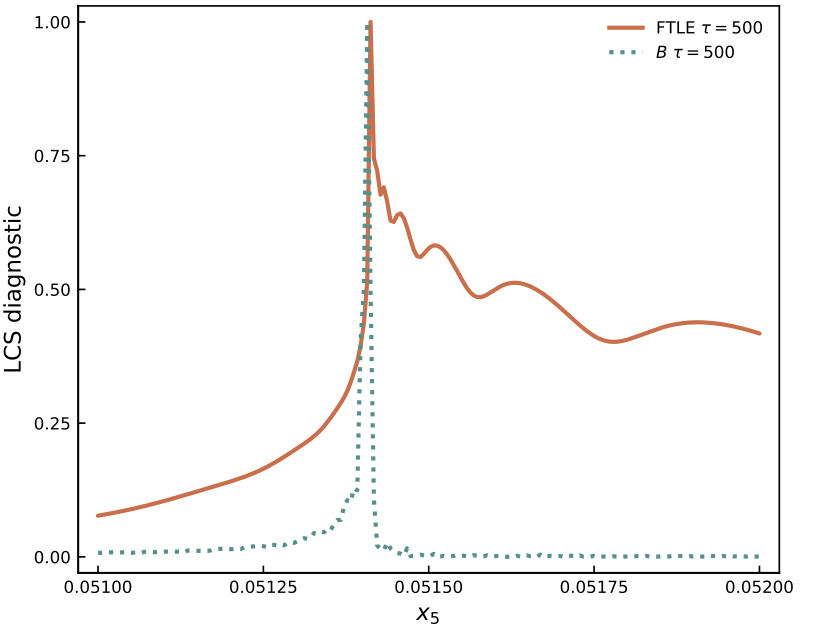}}\put(-230,190){$(c)$} &
        {\includegraphics[width=\columnwidth]{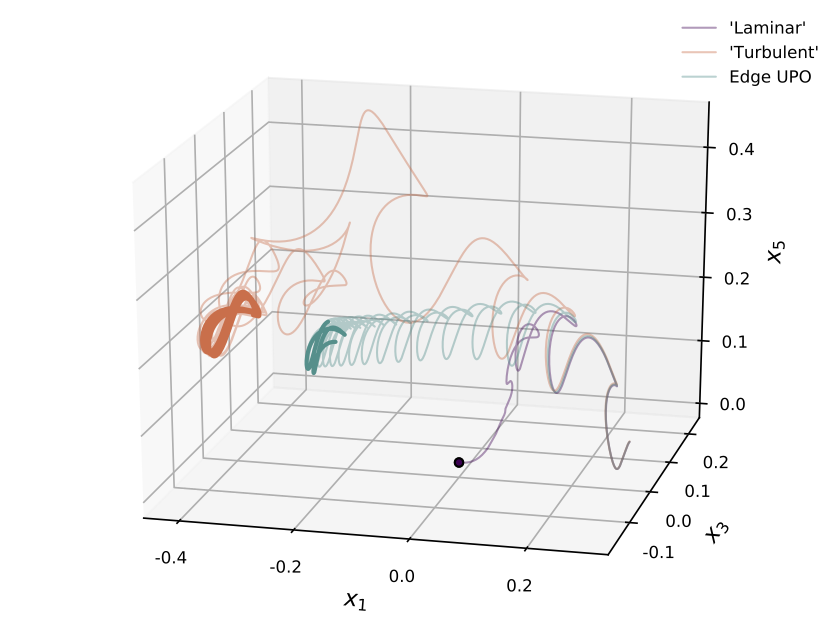}}\put(-230,175){$(d)$}
    \end{tabular}
    \caption{Lagrangian analysis of the LM6D model. $(a)$: FTLE for $\tau=500$. $(b)$: $B$ for $p=0.5$ and $\tau=500$. $(c)$: landscapes of LCS diagnostics from the maps above along a line with constant $x_4=0.4$. Both FTLE and $B$ are normalised with respect to their maximum in the plotted interval. $(d)$: 3D phase portrait of three different trajectories, thicker lines indicate the approached attractor, see text for  details.}
    \label{fig:LM6D_diagnostics}
\end{figure*}
The choice of parameters for the present study shows a bistable system, and thus with a hard edge. The bifurcation diagram in \cite{lebovitz2013edge} shows that for the working parameters there is a stable \emph{laminar} fixed point and a torus to be interpreted as \emph{the turbulent state}. The edge manifold $\Sigma$ separates both basins and contains the lower branch solution originating from a saddle-node bifurcation. This solution is an unstable periodic orbit (UPO) and corresponds to the edge state.
Typical trajectories of LM6D are shown in Fig.\ \ref{fig:LM6D_diagnostics}(d). The Figure shows a phase portrait using three variables $x_1$, $x_3$ and $x_5$. The three plotted trajectories are respectively below $\Sigma$ (purple), above $\Sigma$ (orange) and on $\Sigma$, asymptoting to the UPO (blue) (we assume by default that the manifold $\Sigma$ is orientable).
\begin{figure}
    \centering
    \includegraphics[width=1.00\columnwidth]{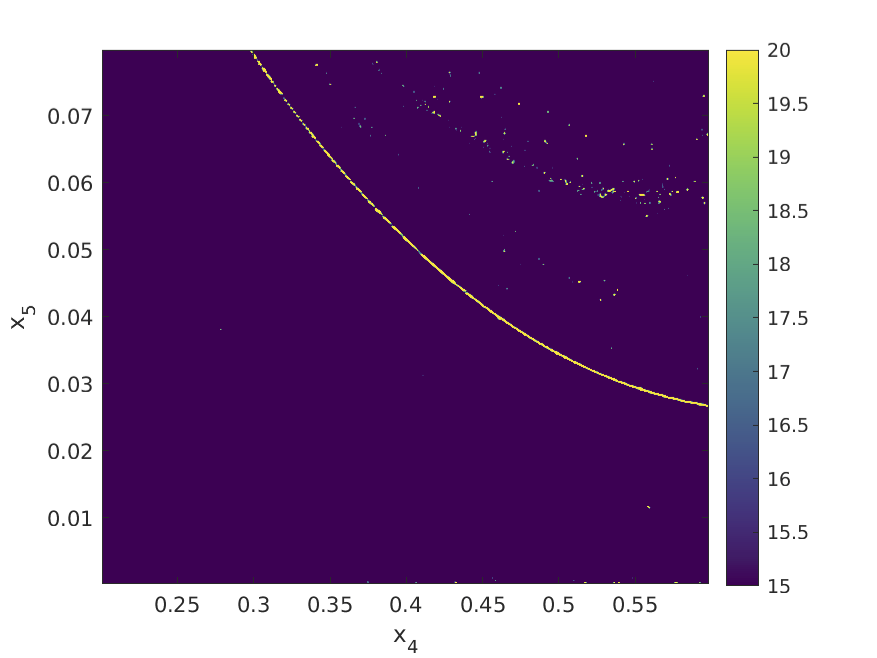}
    \caption{Gradient of the FTLE map for the LM6D model.}
    \label{fig:LM6D_FTLE_grad}
\end{figure}

The LCS diagnostics for LM6D are illustrated in an $x_4 x_5$ plane where the rest of the variables have initial values $x_1=0.3, x_2=x_3=0, x_6=0.1$. The results are shown in Fig.\ \ref{fig:LM6D_diagnostics}(a),(b),(c) for both FTLEs and LDs for a common time horizon $\tau=500$ . A clear ridge is present, consistently with the trajectories approaching different attractors in Fig.\ \ref{fig:LM6D_diagnostics}(d). As for FTLEs, the area on the left of the ridge in Fig.\ \ref{fig:LM6D_diagnostics}(a),(c) is smoother due to the trajectories all approaching the same fixed point, while the right side displays oscillations depending on which part of the torus the trajectory has reached at the final time. In order to enhance the ridge in the FTLE map its gradient is shown in Fig.\ \ref{fig:LM6D_FTLE_grad}. The LD map is computed for $p=0.5$ and shows isovalues of $B$ in Fig.\ $\ref{fig:LM6D_diagnostics}$(b), where a ridge  also emerges, further confirmed in Fig.\ $\ref{fig:LM6D_diagnostics}$(c). These results show that the edge manifold is again highlighted as a repelling LCS in a non-chaotic case when neither the edge state nor the turbulent attractor are fixed points. As we shall see in the next examples this generalisation meets its limits in the presence of chaos.

\subsection{Moehlis-Faisst-Eckhardt model (MFE9D)}

\begin{figure*}
    \centering
    \includegraphics[width=\columnwidth]{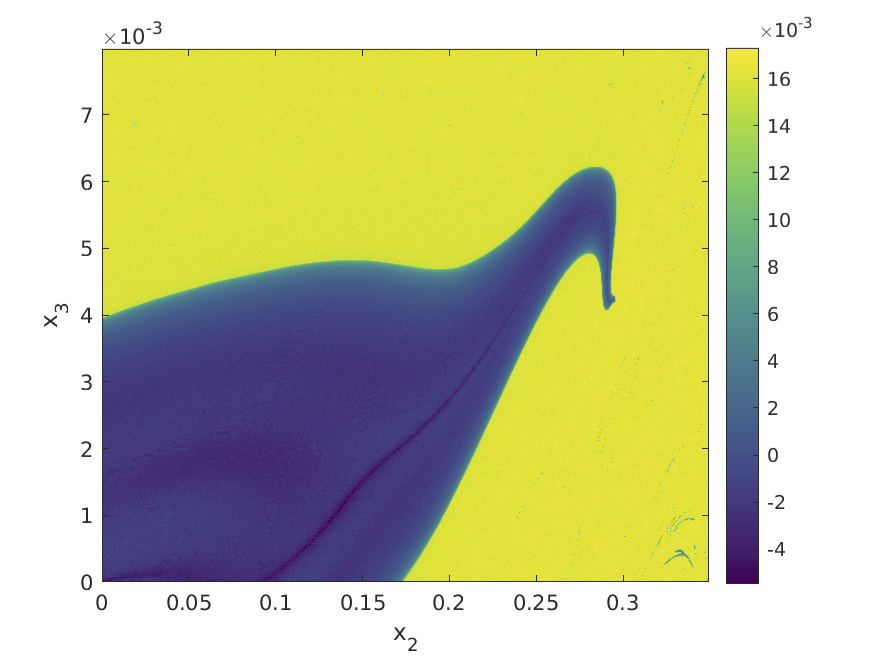}\put(-230,180){$(a)$}
    \includegraphics[width=\columnwidth]{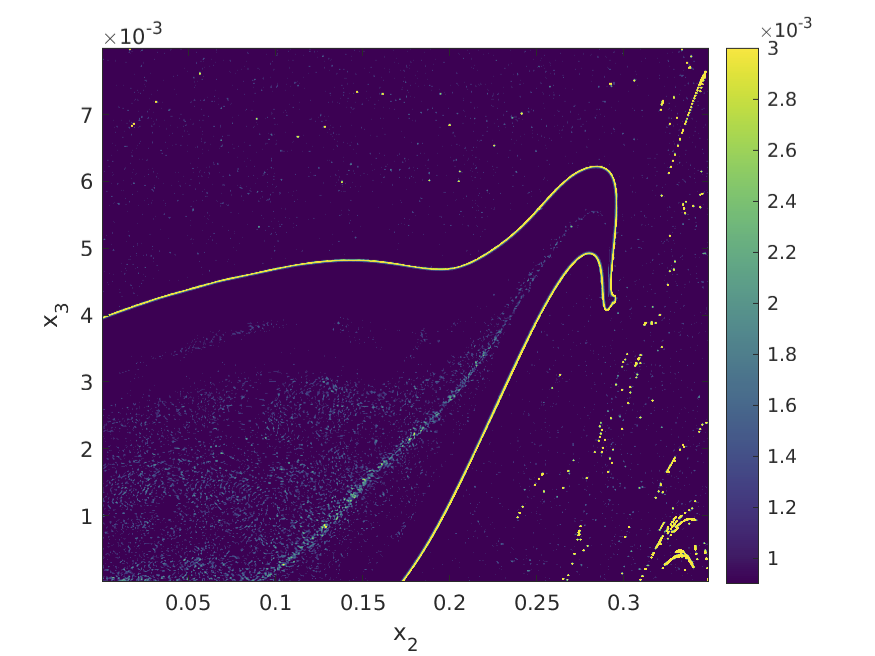}\put(-230,180){$(b)$}\\
    \includegraphics[width=\columnwidth]{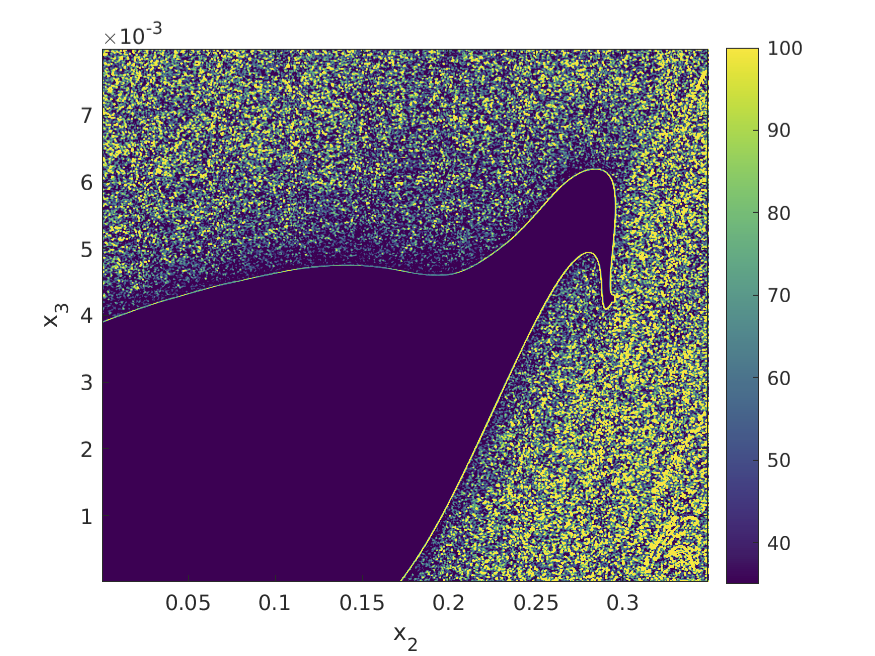}\put(-230,180){$(c)$}
    \includegraphics[width=\columnwidth]{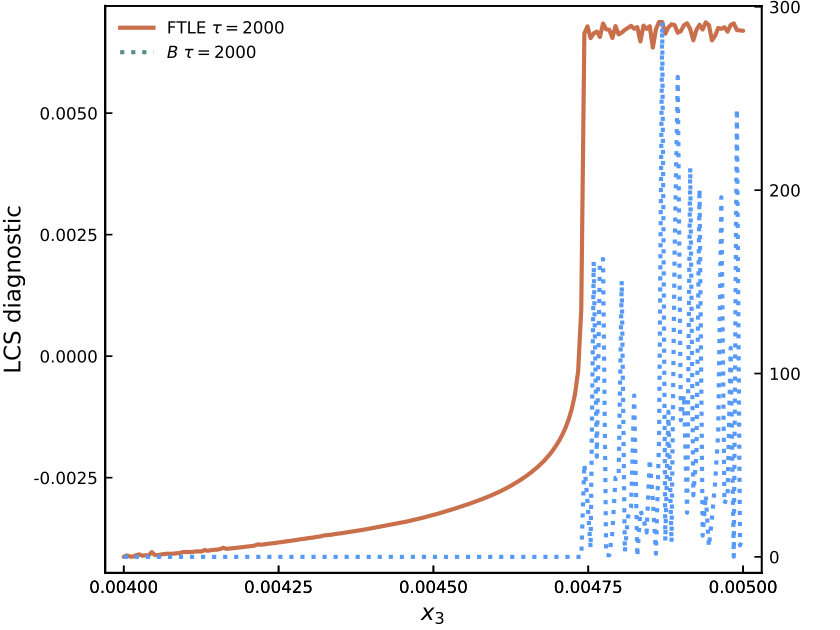}\put(-230,190){$(d)$}
    \caption{Lagrangian analysis of the MFE9D. $(a)$: FTLE for $\tau=800$. $(b)$: norm of the gradient of the FTLE. $(c)$: $B$ for $p=0.5$ and $\tau=800$. $(d)$: landscape of the Lagrangian diagnostics for $\tau=2000$ along a line of $x_3$ in the neighbourhood of the edge for fixed $x_2=0.15$.}
    \label{fig:MFE9D_diagnostics}
\end{figure*}

The next example in the model hierarchy is a nine-dimensional model suggested by Moehlis, Faisst and Eckhardt \cite{moehlis2004}, hereafter referred to as MFE9D. It possesses again a linearly stable laminar state $x_L$, but unlike the models in Section \ref{DM2D} and Section \ref{LM6D} no turbulent-like attractor is present. Instead the turbulent state appears as a chaotic non-attracting set \cite{eckhardt2007turbulence,tel2008chaotic}, the systems posses therefore a weak edge. This model thus contains properties common to real subcritical fluid systems \cite{schmiegel1997fractal, eckhardt2007turbulence}. The parameter defining the state space in the MFE9D is the Reynolds number, set to $\textit{R}=400$ in the current study. For the chosen set of parameters \footnote{The size of the domain is chosen to be $L_X=4\pi$ and $L_Z=2\pi$, fixing the wavenumbers of the Fourier modes of the model.} the dynamics of the model is described in \cite{moehlis2004}, and the edge state corresponds to an UPO \cite{Skufca2006}.

The LCS diagnostics are applied in MFE9D for a horizon time of $\tau=800$. The results are shown in Fig.\ \ref{fig:MFE9D_diagnostics} using a projection on the $x_2x_3$ plane defined by  $x_1=0.7066, x_4=0.01, x_5=x_6=x_7=x_8=x_9=0$. The laminar state lies in the bottom left corner of the figure at $x_2=x_3=0$.

For both LCS diagnostics two different regions, respectively smooth and speckled, emerge in Fig.\ \ref{fig:MFE9D_diagnostics}. They correspond respectively to state space regions with two different behaviours, either uneventful relaminarisation or visiting the chaotic saddle.

The FTLE field is plotted in Fig.\ \ref{fig:MFE9D_diagnostics}(a). It shows a sudden transition between the smooth and speckled regions both in terms of values reached and, unlike the results for the previous models, in terms of fluctuation level. No proper ridge of the FTLE emerges, however by plotting the norm of the gradient of the FTLE with respect to the variables $x_2 x_3$, a ridge stands out in Fig.\ \ref{fig:MFE9D_diagnostics}(b). The performances of the LDs are illustrated for $p=0.5$ in Fig.\ \ref{fig:MFE9D_diagnostics}(c) using isovalues of the gradient norm $B$. In this figure the ridge also stands out around smooth region and separating it from the speckled region. The existence of this sudden transition in the LCS diagnostics is highlighted in Fig.\ \ref{fig:MFE9D_diagnostics}(d) by plotting them along a 1D cut of state space. Fig.\ \ref{fig:MFE9D_diagnostics}(d) shows several abrupt changes of $B$ within the speckled region, however the edge can still be identified as the first abrupt change encountered when coming from the smooth side. Consequently, before using LDs to highlight the edge as an LCS some preliminary knowledge about the qualitative behaviour of the model is required. We attribute again the speckledness property to the presence of chaos in the turbulent basin. Strong visual analogies exist between the results from the LCS diagnostics and the lifetime plots in \cite{Skufca2006} for the same model.

\subsection{Navier-Stokes equations: plane Couette flow (pCf)}
 \begin{figure*}[tb]
    \centering
    \begin{tabular}{cc}
        \includegraphics[width=0.8\columnwidth]{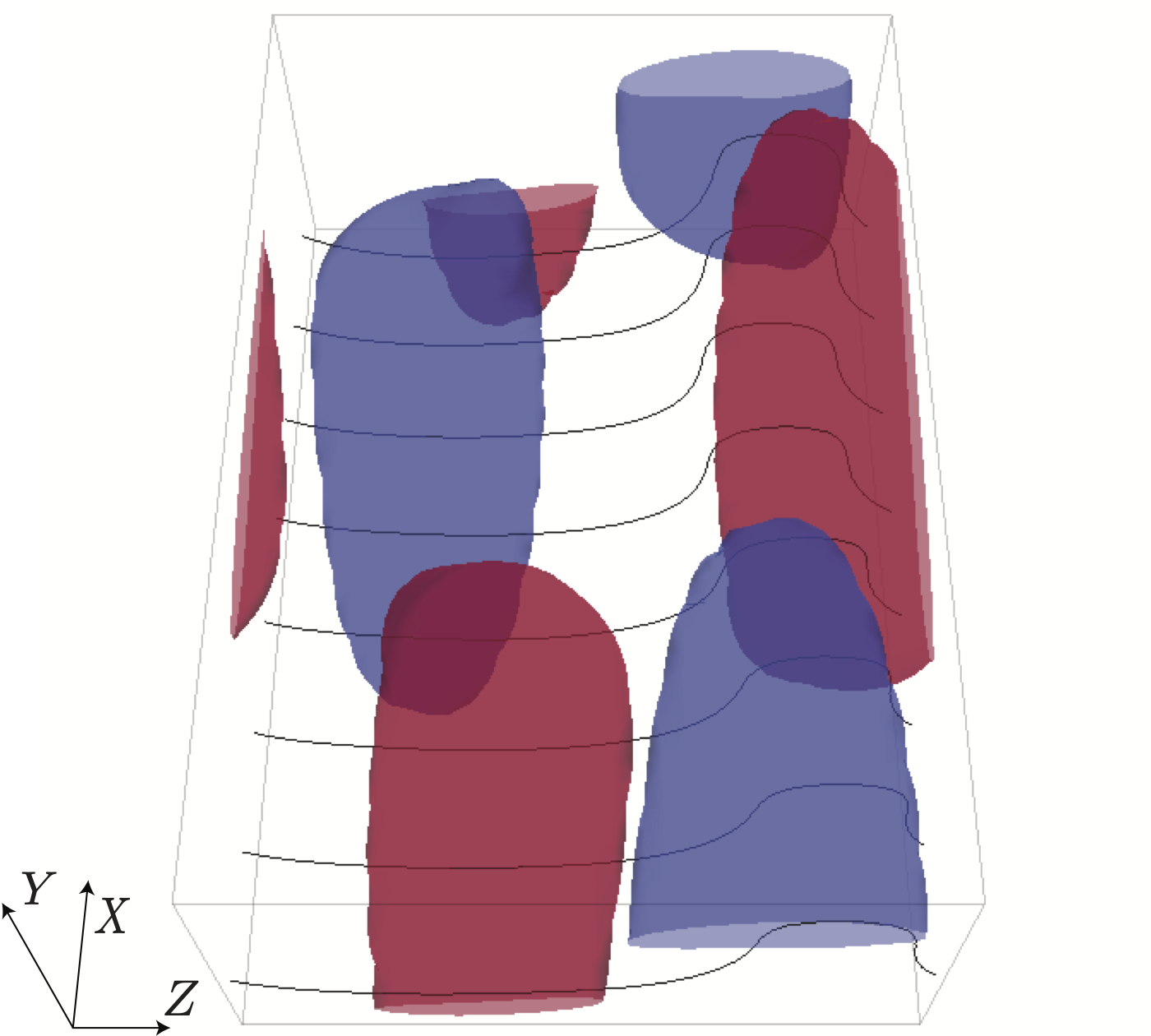}\put(-185,175){$(a)$} &
        \includegraphics[width=0.8\columnwidth]{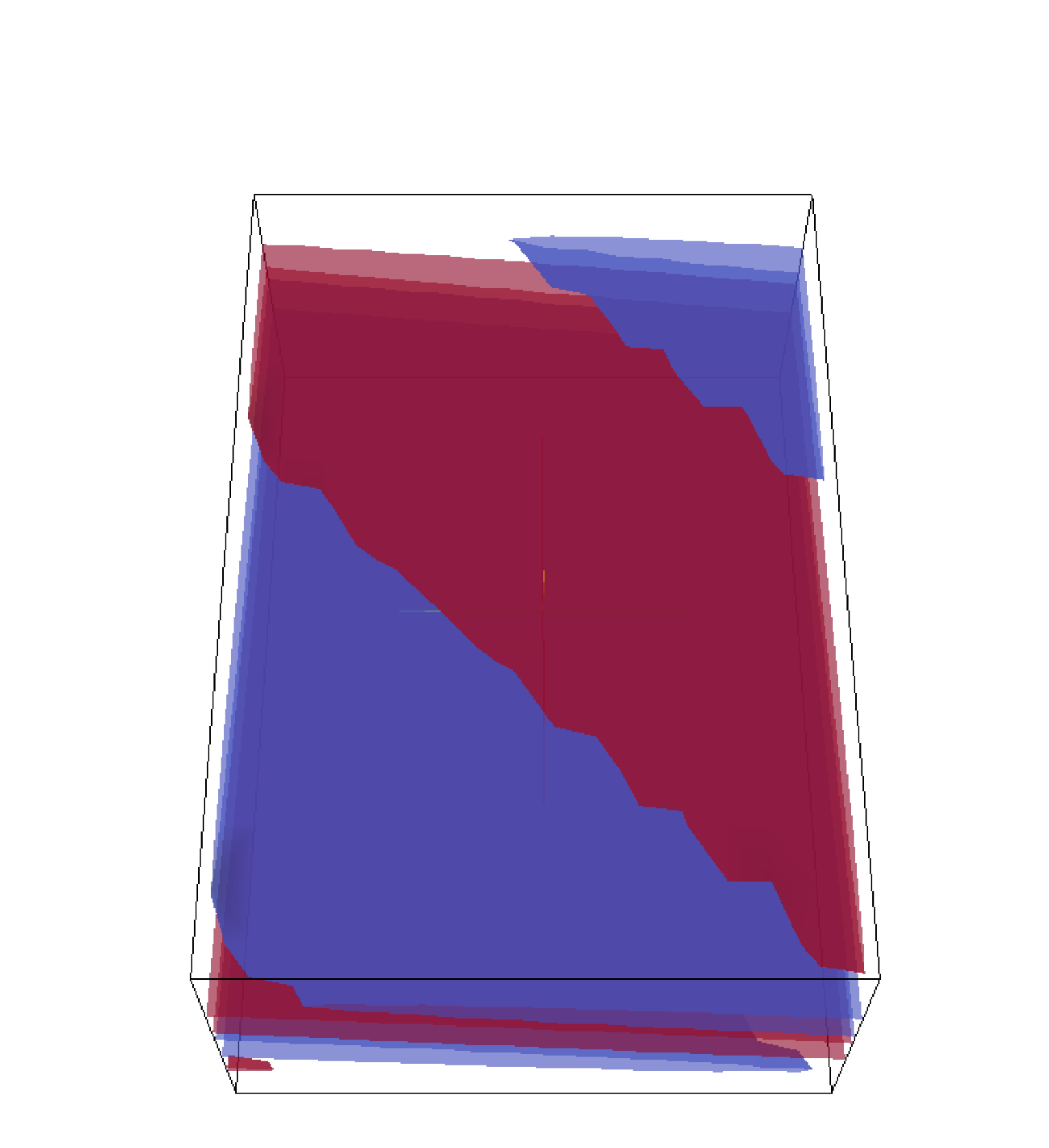}\put(-175,175){$(b)$}
    \end{tabular}
    \caption{Flow fields corresponding to (non-rescaled) initial conditions in Eq.\ \eqref{eq:2Dmap} for plane Couette flow. $(a)$ $\textbf{u}_1$ edge state in \cite{kawahara2001periodic}. Contours of $\mathbf{u}_Z=\pm 0.045$ in red and blue respectively. Black lines: contours of $\mathbf{u}_X=-0.3$ on cross-flow planes at $X=\text{const}$. $(b)$ $\textbf{u}_2$ LOP for wavenumbers $(2\pi/L_X,2\pi/L_Z)$. Contours $\mathbf{u}_Z=\pm 0.45$ in red and blue respectively. Flow along the $x$ direction.}
    \label{fig:pCf_KK_cycle}
\end{figure*}
Eventually we demonstrate the relevance of the previous Lagrangian diagnostics to highlight the edge manifold in a very high-dimensional system governed by the incompressible Navier-Stokes equations. Formally this system is of infinite dimension, however made finite-dimensional by the presence of physical viscosity and by the numerical discretisation. The resulting system is still of such huge dimension -- here of order $\mathcal{O}(10^5)$ -- that scalable methods are necessary for the diagnostics. We focus here on plane Couette flow (pCf), where a viscous fluid is sheared between two plates moving with opposite velocities in a direction called $X$. pCf is parametrised by a non-dimensional Reynolds number $R$ proportional to the plate velocities. All the quantites are made non-dimensional with the half gap between the plates $h$ and the plate velocity $U_w$, the Reynolds number is $R=U_w h/\nu=400$, with $\nu$ the kinematic viscosity. The Navier-Stokes equations for the perturbation velocity field to the laminar state ${\bm U_L}$, $\mathbf{u}=(u_X,u_Y,u_Z)$, read
\begin{multline}
     \frac{\partial \mathbf{u}}{\partial t} +(\mathbf{u}\cdot \nabla) \mathbf{u} + (\mathbf{u}\cdot \nabla) \mathbf{U_L} + (\mathbf{U_L}\cdot \nabla) \mathbf{u} = \\ -\nabla p + R^{-1}\nabla^2 \mathbf{u} \label{eq:Navier_Stokes}
\end{multline}
and
\begin{equation}
    \nabla \cdot \mathbf{u} = 0, \label{eq:continuity}
\end{equation}
where $p=p(X,Y,Z,t)$ is the pressure field \footnote{Numerical simulation of the Couette system is performed using the well-tested in-house unsteady spectral code SIMSON \cite{chevalier2007simson}. Periodic boundary conditions are imposed in the $X$ and $Z$ directions, while Dirichlet boundary conditions apply at the walls at $Y=\pm$1. The resolution used is the same as in \cite{kawahara2001periodic}, where a very similar spectral decomposition is used.}.
pCf is the archetype of a subcritical system, where for all values of $R$ large enough, a turbulent state $T$ coexists with a linearly stable laminar state $L$. The laminar state has a velocity field ${\bm U_L}=y\bf{e_X}$, where $\bf{e_X}$ is the longitudinal unit vector, and a homogeneous pressure field. More precisely, we consider as in Ref. \cite{kawahara2001periodic,kawahara2005} that the fluid particles move inside a numerical domain of dimensions $L_X \times  L_Y \times L_Z$ (in units of $h$), made artificially periodic in both in-plane directions $X$ and $Z$ and bounded in the wall-normal direction $Y$. The exact dynamics of the edge state depends on the parameters $R$, $L_X$, $L_Y$ and $L_Z$. However there are robust common features to all of them: the presence of streamwise streaks (transverse spatial modulations of the streamwise velocity field, weakly modulated in the longitudinal direction, associated with longitudinal vortices). For the parameters under study ($R$=400, $(L_X,L_Y,L_Z)=(5.513,2,3.770)$) the edge state is known and consists of a time-periodic flow field with period $T_p=$85.5\cite{kawahara2001periodic}. Its spatial structure is illustrated in Fig.\ \ref{fig:pCf_KK_cycle}(a). The structure of its stable manifold, however, and of the edge manifold in general, is however poorly understood despite some recent progresses \cite{Kreilos2012, chantry2014studying}. For this parameters, as in the MFE9D model, the turbulent trajectories are supertransients \cite{tel2008chaotic} and there is no turbulent attractor.

An observable characterising the laminar-turbulent transition is is the streamwise vorticity squared $|\omega_X|^2$ averaged over the computational domain $a(t)= (1/V\int|\omega_X|^2dv)^{1/2}$, where $\omega_X=\partial_Y u_Z - \partial_Z u_Y$ and $V$ is the volume of the computational domain. In order to identify the edge manifold $\Sigma$ as an LCS, we use an arbitrary point on the edge state as a reference. That state space point corresponds to a flow field in the three-dimensional physical space $(X,Y,Z)$. We will use the perturbation velocity field notation $\bm{u_{E}}$ for such a point. Note that $\bm{u_{E}}$ is the perturbation to $\bf{U_L}$ and the full velocity field is $\bf{V}=\bf{U_L}+\lambda \bm{u_{E}}$ with $\lambda$=1 corresponding to the edge state. By tuning a real parameter $\lambda$, one expects a change in behaviour of both FTLEs and LDs at $\lambda$=1.
In order to extend the definition \eqref{eq:Mfull} to the Navier-Stokes case, we first need to specify which projection system is used to define the corresponding high-dimensional state space. The choice is not unique. For simplicity we choose to rely on the values $u_k({\bm X}_i)$ of the velocity at each discrete grid point, where the index ($k=$1,2,3) refers to the velocity component ($k=X,Y,Z$) and $i=1,...,N$, where $N=N_X \times N_Y \times N_Z$ is the total number of grid points.

The expression for the Lagrangian Descriptor scalar valued $M^+$ along a trajectory is
\begin{equation}
      M^+(\mathbf{V_0},t_0,\tau)=\int_{t_0}^{t_0+\tau} \sum_{k=1}^{3}\sum_{i=1}^{N}\left| \frac{\partial \mathbf{V}_{k}({\bm X}_i)}{\partial t}(t;\mathbf{V_0}) \right|^{p} dt,\label{eq:M_NS}
\end{equation}
where $\tau$>0.
Other physically meaningful formulations are possible, e.g.\ a velocity-vorticity formulation and/or a spectral Galerkin decomposition.  If we denote by $\gamma$ the state space coordinate along which the bisection is performed, the gradient of $M^+$ in that direction is given by $\partial_{\gamma} M^+(\mathbf{V_0}(\lambda),t_0,\tau)$.
If $B$ is evaluated in a $d$-dimensional region of the state space (with $d \ll N$), its partial derivatives need to be evaluated only along these $d$ directions, each direction being parametrised by a coordinate $\gamma_i$. The expression for $B$ becomes
\begin{equation}
       B = \left(\sum_{i=1}^{d}\left(\frac{\partial M^+(\mathbf{V_0}(\lambda),t_0,\tau)}{\partial \gamma_i}\right)^2\right)^{1/2}.
\end{equation}

$B$ is computed numerically using second order centered finite differences as in Eq.\ \eqref{eq:discreteB}. In all simulations $\varepsilon=10^{-9}$ and $p$=0.5.

The direct computation of all FTLEs using classical methods is indubitably out of reach in the present system because of the huge value of $N$. The so-called reduced-order FTLEs can however be accessed as a by-product of the calculation of Optimally Time-Dependent (OTD) modes \cite{babaee2016minimization}. In this approach, a finite number $r$ of these time-dependent vectors can be evolved in time together with the main trajectory.
\begin{figure}
    \centering
    \includegraphics[width=0.45\textwidth]{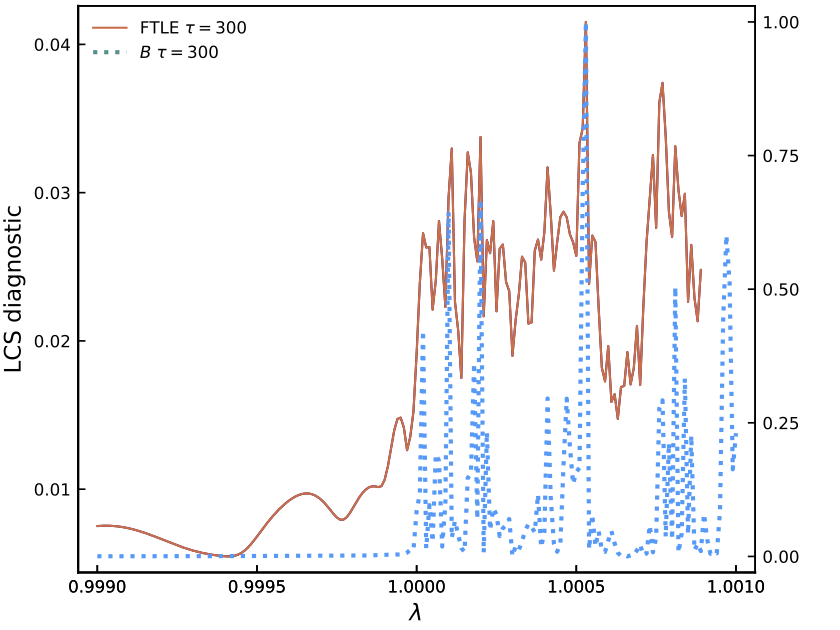}
    \caption{One-dimensional landscape of Lagrangian diagnostics for several values of  $\lambda$ in the neighbourhood of the edge at $\lambda=1$. Plane Couette flow, same parameters as \cite{kawahara2001periodic}.}
    \label{fig:KK_diagnostics}
\end{figure}
They have the property of spanning a reduced $r$-dimensional space which approximates the tangent space, from which the finite-time exponents can be directly computed under mild conditions \cite{babaee2017,feppon2019extrinsic}.
\begin{figure*}
    \centering
    \includegraphics[width=0.49\textwidth]{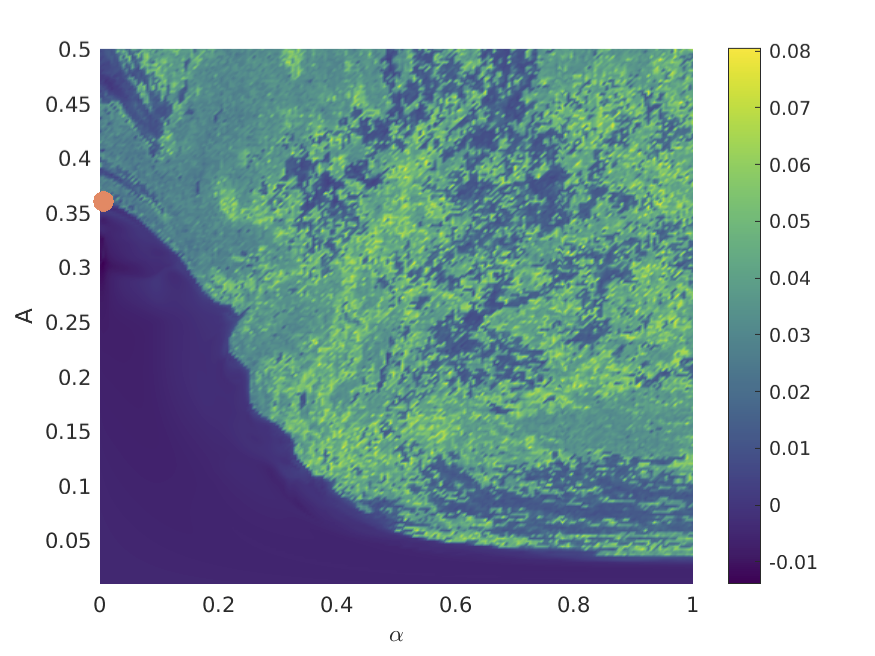}\put(-245,180){$(a)$}
    \includegraphics[width=0.49\textwidth]{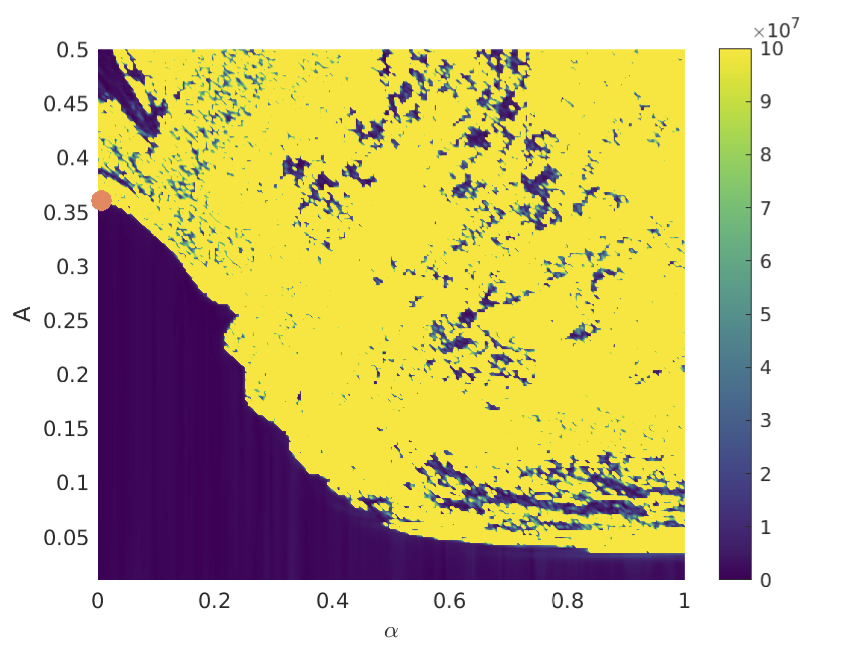}\put(-245,180){$(b)$}
    \caption{2D maps of LCS diagnostics for plane Couette flow, same parameters as Ref. \cite{kawahara2001periodic}, $\tau=300$. Left: largest reduced-order FTLE. Right: isovalues of $B$ for $p=0.5$. The boundary between the speckled and the smooth area corresponds to the finite time approximation of $\Sigma$. The orange dot indicates a projection of the edge state in the $\alpha$-$A$ state portrait.}
    \label{fig:KK_2D_map}
\end{figure*}
Although only the leading FTLE is of interest here, it is recommended to compute them with $r$>1 to avoid spurious results linked with eigenvalue crossings \cite{babaee2017}. The present simulations involve $r=4$ OTD modes. The four corresponding initial conditions are the edge state point $\bm{u_{E}}$, and the three linear optimal perturbations (LOPs) computed for the wavenumbers $(k_X,k_Z)$=$(0,2\pi/L_z)$, $(2\pi/L_x,0)$ and $(2\pi/L_x,2\pi/L_z)$. For each wavenumber $(k_X,k_Z)$, the corresponding LOP is defined in \cite{schmid2001stability}
as the initial condition ($\mathbf{u_0})$ on the unit sphere maximising the energy gain $G_{\tau}$, where
\begin{equation}
       G_{\tau}=\left|\left| F_{t_0}^{t_0+\tau}(\mathbf{u_0})\right|\right|^2,
\end{equation}
where $\left|\left|\cdot\right|\right|$ is the usual $L^2$-norm and $F_{t_0}^{t_0+\tau}$ is the propagator associated with the Navier--Stokes operator \cite{schmid2001stability} linearised around $\bf{U_L}$, which is independent of $t_0$.
The value of $\tau$ used here is the one that gives the largest value of $G_\tau$ over all initial conditions. Interestingly the quantity $\left(\log{G_\tau}\right)/2\tau$ can be interpreted as the leading FTLE at time $t_0$ over a time horizon $\tau$ around the fixed point $\bf{U_L}$ at time $t_0$, in the subspace spanned by the corresponding eigenvector \cite{cossu2010secondary}. It is known that eigenvalue crossing can affect the resulting OTD subspace \cite{babaee2017} and therefore the reduced-order FTLEs. In order to converge to the relevant subspace, the OTD modes are first evolved for $200$ time units along the edge trajectory and then used as an initial condition for the computation of the reduced-order FTLEs.

In Fig.\ \ref{fig:KK_diagnostics} and \ref{fig:KK_2D_map} we show landscapes of the two indicators in a one-dimensional cut and in a two-dimensional slice of the state space, respectively. Fig.\ \ref{fig:KK_diagnostics} displays the 1D landscape of LCS diagnostics as $\lambda$ is varied, for a time horizon $\tau=300$, for both the leading FTLE and $B$.
The FTLEs display a smooth behaviour for $\lambda < 1$, where trajectories approach the laminar state, while for $\lambda > 1$ the landscape looks jagged in a way similar to the MFE9D in Fig.\ \ref{fig:MFE9D_diagnostics}(d). The LD landscape also displays the same properties, however the values of $B$ in laminar basin are considerably lower than for FTLEs, resulting in a flatter landscape. In both cases the crossing of the edge manifold at $\lambda$=1 corresponds to a steep increase of the quantity plotted, this property being better visible in the case of $B$. This steepness property has its analog in lifetimes studies in former works dedicated to chaotic saddles \cite{bollt2005path,Skufca2006,munoz2012}.
The geometry of the state space can be further explored by considering a two-dimensional slice spanned by two perturbation fields $\mathbf{u}_1$ and $\mathbf{u}_2$. These two fields are normalised so that $\norm{\mathbf{u}_1}=\norm{\mathbf{u}_2}=1$. An initial perturbation $\bf{u_0}$ to the laminar state can be defined as
\begin{equation}
    \mathbf{u_0}(\alpha,A)= A\frac{(1-\alpha) \mathbf{u}_1+\alpha\mathbf{u}_2}{\left|\left|(1-\alpha) \mathbf{u}_1+\alpha \mathbf{u}_2\right|\right|} \label{eq:2Dmap}
\end{equation}
where $A$ is interpreted as an amplitude. We chose $\mathbf{u}_1=\mathbf{u}_{E}$, such that for $\alpha=0$ the same subset of state space is explored as in \ref{fig:KK_diagnostics}(a). As for  $\mathbf{u}_2$ we chose the LOP corresponding to the wavenumbers $(k_X,k_Z)=(1,1)$.
The velocity fields for the initial conditions $\mathbf{u}_i\  i=1,2$ are shown in Fig.\ \ref{fig:pCf_KK_cycle}. The resulting LCS diagnostics for $\tau=300$ are shown in Fig.\ \ref{fig:KK_2D_map}. In each subfigure, two very different regions appear, separated by a smooth boundary. It has been monitored that the zone containing the laminar state corresponds to rapidly relaminarising trajectories, whereas the lighter zone on the other side of the boundary contains trajectories visiting the turbulent state. In other words, for both diagnostics the basin boundary is successfully identified in finite time.
For both indicators, some 'holes' appear inside the turbulent basin, characterised by low values typical of the laminar basin. No strong difference between FTLEs and LDs appears in the comparison between the two subfigures. These holes correspond to initial conditions which relaminarise after a turbulent transient shorter than $\tau$. A qualitative comparison with lifetime maps as used in \cite{Skufca2006} and \cite{Kreilos2012} suggests again similar features.

\section{Edge tracking revisited}
\label{edgetracking}
The previous sections have demonstrated that LCS diagnostics are efficient to capture stable manifolds in their globality (the edge), we now show that these diagnostics can be adapted in order to identify the relative attractors sitting on them, i.e.\ edge states. In dissipative systems of interest, edge states are typically invariant sets of low dimension e.g.\ fixed points, limit cycles or low-dimensional chaotic sets. In the hydrodynamic literature, edge states have proven crucial to unfold bifurcation diagrams \cite{Kreilos2012, avila2013streamwise}, for dynamic control \cite{kawahara2005} and because of their role in turbulence nucleation \cite{Khapko2016, kreilos2016bypass}. Until now edge states have been routinely identified numerically using standard bisection methods coupled with prior knowledge of the location of the edge states \cite{itano2001dynamics,Skufca2006}. The present section introduces a new class of bisection methods based on LCS diagnostics.  We will contrast the related concepts of global vs. local bisection methods in terms of their ability to identify the edge state, their scalability and their associated computational cost.

\subsection{Global vs. local methods}

The standard bisection method used for edge tracking relies on some preliminary knowledge of the part of state space where the edge state resides. A scalar observable $a(t)$ (typically the perturbation kinetic energy) is chosen such that the edge state lies entirely in an interval $a \in (\alpha_A,\alpha_B)$ whereas the other states (L and T for simplicity) lie outside it. Provided this constraint is met, the initial condition $x_0$ is adjusted recursively so that $F_{t_0}^{t_0+\tau}(x_0)$ remains in $(\alpha_A,\alpha_B)$ for $\tau$ as large as possible. Such a method warrants that $F_{t_0}^{t_0+\tau}(x_0)$ converges towards the edge state as $\tau \rightarrow \infty$. We label this first bisection method as \emph{global}, since it uses information from the whole state space to define $\Sigma$. It has proven useful in the presence of both a hard and a soft edge manifold.

Global methods can however fall short whenever no available global observable can elucidate on which side of the edge the different trajectories evolve \cite{zammert2019transition,beneitez2019edge}, when one of the attractors undergoes a local bifurcation affecting the values of the bounds \cite{canton2020critical} or simply when information on the bounds $\alpha_{A,B}$ is not available.

As shown in the examples above, the LCS diagnostics are based on the knowledge of the tangent vector and the Cauchy-Green tensor, which are local in state space and rely only on a finite-time description. Since they are sufficient to highlight the edge manifold, we expect these diagnostics to yield revised \emph{local} bisection methods by opposition to the \emph{global} ones. The use of a local method for bisection offers the following advantages:
\begin{itemize}
    \item No preliminary choice of observable is needed based on physical intuition
    \item No prior knowledge of bounds $\alpha_{A}$ and $\alpha_{B}$ is needed
    \item The approach to edge tracking is not binary, edge trajectories are labelled as optimisers of an LCS-based observable.
    \item The edge tracking has a predefined time horizon $\tau$ chosen as an intrinsic parameter. The values of $\tau$ are limited by numerical accuracy, however local edge tracking can be restarted from any location, making the algorithm iterative and eventually convergent.
    \item LCS-based methods allow for a quick identification of the regions of interest.
\end{itemize}
Although the distinction between local and global methods is pedagogically convenient, the designation \textquote{local method} needs however to be nuanced in practice: the shorter the time horizon $\tau$ the more local the approach. For larger values of $\tau$, non-convergent trajectories are not bound to local neighbourhoods and can explore remote parts of the state space. The analysis of the LCS diagnostics in the previous section suggests long time horizons $\tau$ for a sharper idenfication of the edge.

\subsection{Comparison of different methods for edge tracking}
\label{edgetracking_difmeth}
\begin{figure}
    \centering
    \includegraphics[width=0.75\columnwidth]{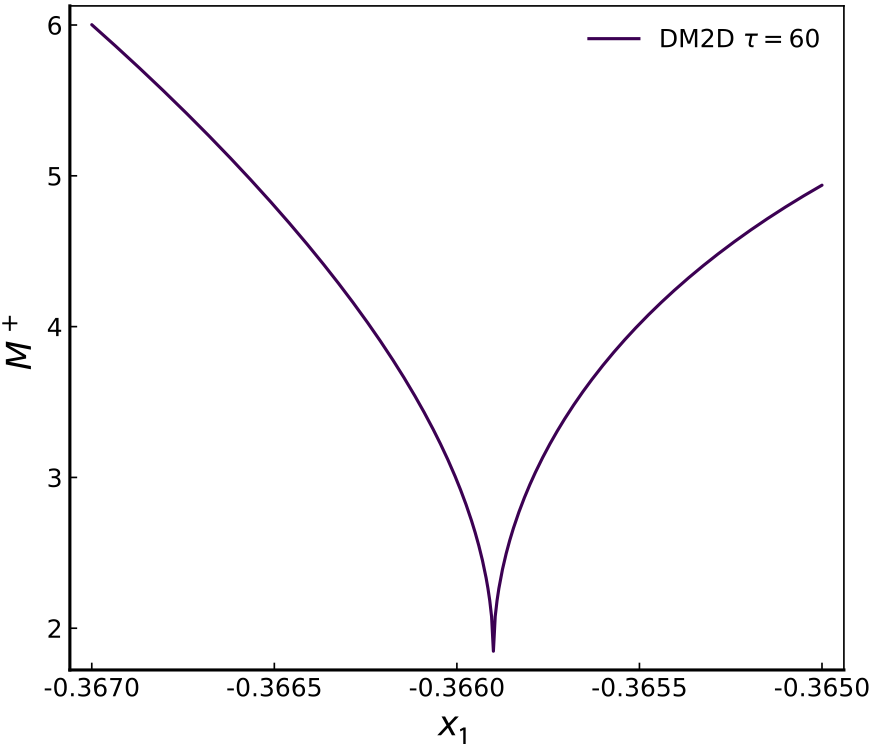}\put(-185,165){$(a)$} \\
    \includegraphics[width=0.75\columnwidth]{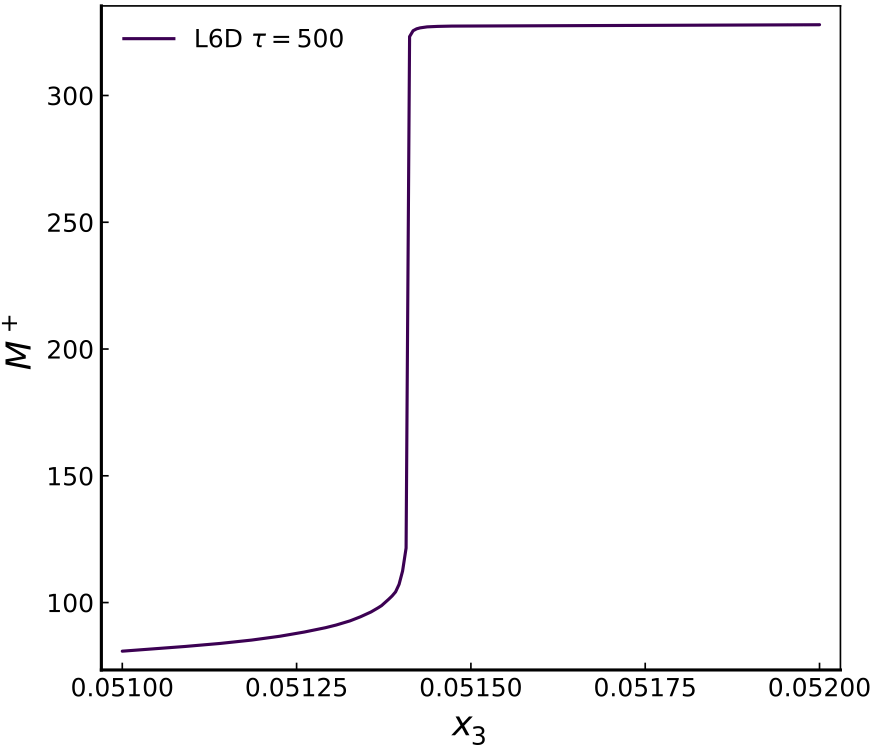}\put(-185,155){$(b)$}\\
    \includegraphics[width=0.75\columnwidth]{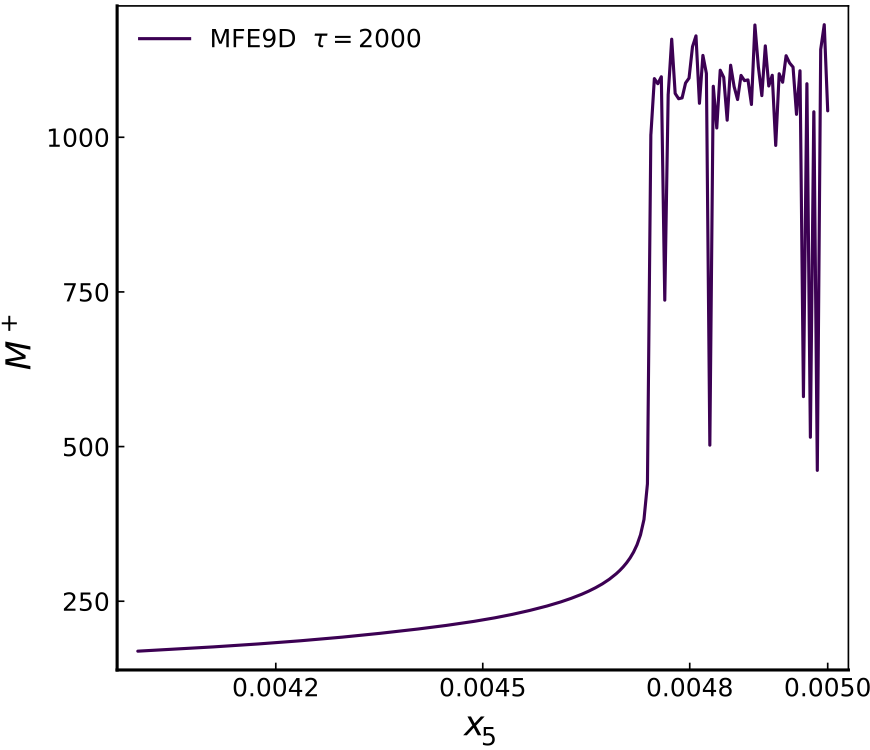}\put(-185,155){$(c)$}
    \caption{One-dimensional landscapes of the forward-time Lagrangian Descriptor $M^+$ (with $p=$0.5) for the models (a) DM2D (edge state: fixed point, turbulent state: attracting fixed point) (b) LM6D (edge state: periodic orbit, turbulent state: attracting 2-torus) and (c) MFE9D (edge state:  periodic orbit, turbulent state: chaotic saddle).}
    \label{fig:LD_landscapeComp}
\end{figure}

\begin{figure*}
    \centering
    \includegraphics[width=0.49\textwidth]{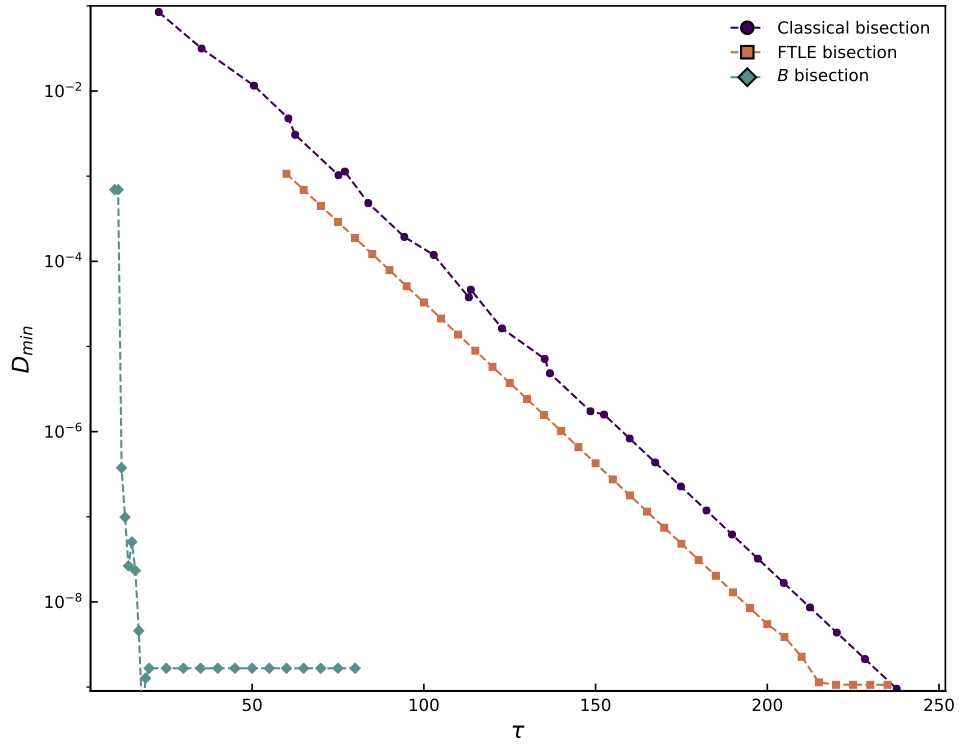}\put(-245,195){$(a)$}
    \includegraphics[width=0.49\textwidth]{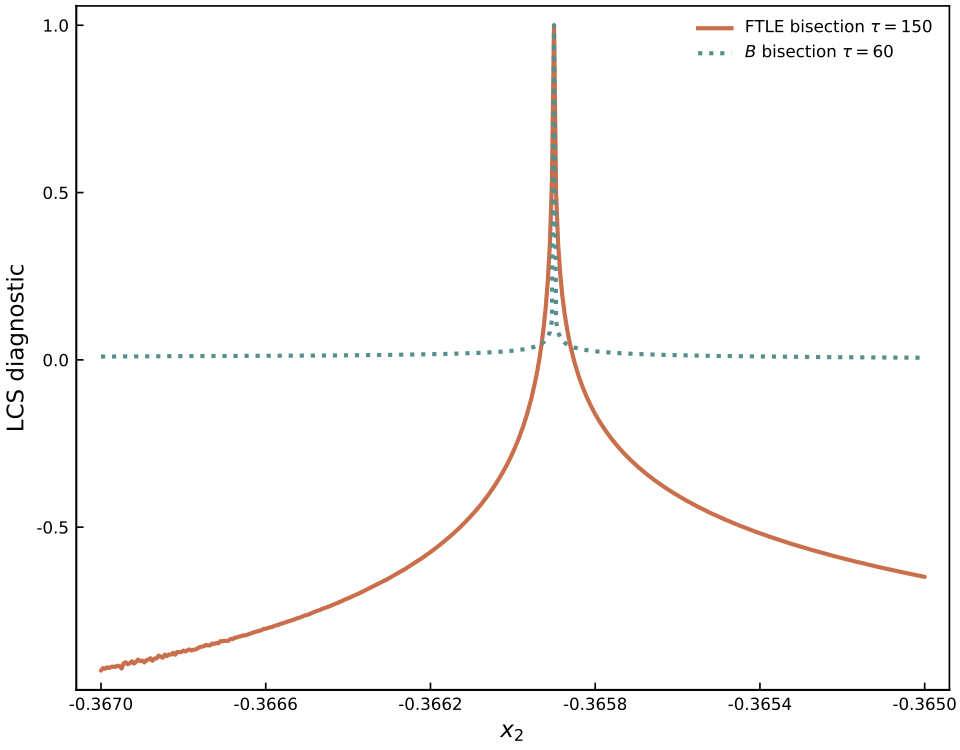}\put(-245,195){$(b)$}
    \caption{Results for edge tracking along $x_2$ in the DM2D model, for a fixed $x_1=0$. Left: Euclidian distance to the edge state $x_E$ depending on the objective time $\tau$ of the bisection algorithm employed. Tolerance for algorithm convergence $10^{-9}$. Right: LCS diagnostic landscape for $x_2$ in the neighbourhood of $\Sigma$. Both FTLE and $B$ are normalised with respect to their maximum in the plotted interval.}
    \label{fig:DM2D_bisectionComparison}
\end{figure*}

LCS-based bisection relies as usual on the iterative process of straddling the edge trajectory locating the right initial condition along an arbitrary state space direction. Unlike with the classical bounds-based method, the revised edge trajectory emerges now as the optimiser of a given functional rather than from the difficultly quantifiable \textquote{neither laminar, nor turbulent} definition. For a given time horizon $\tau$>0, FTLE-based bisection seeks the initial condition $x_0$ maximising $\lambda_{t_0}^{t_0+\tau}(x_0)$. LD-based bisection seeks either the minimum of $M^+(x_0)$ or the maximum of the gradient norm $B(x_0)$, depending on the dynamical nature of the edge state. The various examples in Section \ref{modelsSection} point towards the following phenomenology: when the edge state is a fixed point it is sufficient to use minima of $M^+$, however this criterion needs to be replaced by maxima of $B$, which is computationally more costly, for any edge state dynamics of higher dimensionality. These different cases are contrasted in Fig.\ \ref{fig:LD_landscapeComp} where one-dimensional landscapes of $M^+$ are displayed. FTLE landscapes are shown for comparison in Fig.\ \ref{fig:DM2D_bisectionComparison}(b), \ref{fig:LM6D_diagnostics}(c) and \ref{fig:MFE9D_diagnostics}(c).

The simplest case for the comparison of the different edge tracking methods is the two-dimensional Dauchot-Manneville model (DM2D), for which the edge state $x_E$ there has an analytical expression in Eq.\ \eqref{eq:xDM2D}. For the sake of generality, the LD-based edge tracking is not defined using $M^+$ but rather using its gradient $B$, whereas the FTLE-based method is based on the estimation of the largest FTLE. The present methods rely, for the proof of concept, on simple algorithms to locate the maximum  of the associated fields along an arbitrary one-dimensional line $\mathcal{L}$ in state space (in practice the line $\mathcal{L}:\{x_1=0\}$ was selected). The maximisation is always initiated on the \textquote{smooth} side, i.e.\ within the basin of attraction of $x_L$.

The maximisation over the line $\mathcal{L}$ of the LCS diagnostic ($\lambda^{\tau}(x_0)$ or $B(x_0)$) generates a sequence of new initial conditions $x_0^{(k)}$, $k$=0,1,2,... on the line $\mathcal{L}$. Convergence is satisfied if the sequence $x_0^{(k)}$ approaches asymptotically to some $x_0^* \in \Sigma$. Since the goal is to identify $x_E$, a convergence distance $D_{min}$, based on the minimal Euclidian distance along the resulting trajectory to $x_E$, is preferred. It is defined as
\begin{equation}
D_{min}(x_0^*,\tau) = \min_{t \in (t_0,t_0+\tau]} \left|\left|F_{t_0}^t(x_0^{*})-x_E\right|\right|. \label{eq:Dmink}
\end{equation}
where the minimum is taken over the time interval $(t_0,t_0+\tau ]$. The largest FTLE is computed using finite differences with $\varepsilon=10^{-9}$. As for the Lagrangian Descriptors, the gradient $B=\left|\left|\partial M^+/\partial x_{0,2}\right|\right|$ is computed following Eq.\ \eqref{eq:discreteB} using $p=0.5$.

Fig.\ \ref{fig:DM2D_bisectionComparison}(a) shows the minimum distance $D_{min}$ to the edge state for the classical, the FTLE-based and the LD-based edge tracking algorithms. In the classical edge tracking, $\tau$ >0 is the time it takes for a single trajectory closest to the edge to reach the bounds, and it varies from iteration to iteration. In the local edge tracking algorithms however, $\tau$>0 is a prescribed time horizon. Fig.\ \ref{fig:DM2D_bisectionComparison}(a) shows that, for equivalent $D_{min}$, the so-called local methods both require a lower value of $\tau$ compared to the classical method.

A direct mutual comparison of the local methods suggests that FTLE-based edge tracking requires time horizons $\tau$ one order of magnitude larger than the LD-based edge tracking to achieve equivalent $D_{min}$. Fig.\ \ref{fig:DM2D_bisectionComparison}(a) also illustrates the bottom value reached by $D_{min} \approx 10^{-9}$. This can be understood as follows: FTLEs, unlike LDs, do not display any ridge near hyperbolic saddle points if the dynamics is governed by a linear system \cite{HALLER2011574,lopesino2017theoretical}. As a consequence, the trajectory in the neighbourhood of the saddle point $x_E$ must \textquote{feel} nonlinear effects for FTLEs to be able to identify the edge as a LCS. This implies that longer times are needed for the bisection based on FTLEs compared to that based on LDs. Similarly convergence is expected to be faster in the LD case, at least when the edge state corresponds to a fixed point. Longer times can be requested in cases with more complex attractors, although this demands a more systematic investigation.

The convergence properties of these two LCS-based edge tracking algorithms in higher dimension are expected to proceed along the same lines because of the strong \emph{scalability} properties of the diagnostics themselves.  In the case of LDs, the function $M^+$ is computed directly along the trajectories, without using the Jacobian operator $\nabla_x f(x(t),t)$. The function $M^+$ is hence trivially generalised to arbitrary dimension according to \eqref{eq:Mfull}, as demonstrated in \eqref{eq:M_NS}. The full computation of all FTLEs, on the other hand, becomes unfeasible in higher dimension since $\nabla F_{t_0}^{t}$ is of prohibitive size $n \times n$, as pointed out in \cite{babaee2017}. The computation of $r$ OTD modes, with $1 \le r\ll n$, can however grant access to the $r$ leading FTLEs at a cost $n\times (r+1)+r^2$ \cite{babaee2017}. For $r$=1 this amounts to $\mathcal{O} (2n)$, the computation of the largest FTLE is hence based on a scalable algorithm as well.

\section{Conclusion, discussion and outlooks}

Motivated by the physical problem of subcritical laminar-turbulent transition to turbulence in hydrodynamics, we have investigated the established notion of edge manifold which charts the state space into different basins of attraction \cite{eckhardt2007turbulence}. The main contribution of this study is the recognition of stable manifolds of edge states in autonomous systems of arbitrarily high dimensions as Lagrangian Coherent Structures (LCSs). The edge manifold divides the state space in the same way as a hyperbolic repelling Lagrangian Coherent Structure, and hence a direct relation between LCS and exact coherent states can be established. The same mathematical toolbox used to highlight an LCS can then be used to highlight the edge manifold, no matter how high the dimension of the underlying state space. The price to pay for this generalisation is that only forward-time operators can be considered, which prevents one from considering attracting LCS sets such as unstable manifolds. An illustration based on two LCS diagnostics, Finite-time Lyapunov exponents (FTLEs) \cite{Haller_2015} and forward-time Lagrangian Descriptors (LDs) \cite{mancho2013lagrangian}, has revealed the edge manifold in several autonomous dynamical systems of increasing dimension and complexity. By order of dimension, these models are: i) a didactic bistable two-dimensional model with only fixed points \cite{Dauchot_1997}, ii) a bistable six-dimensional model where the edge state is an unstable periodic orbit and turbulence corresponds to a stable periodic orbit \cite{mariotti2011low}, iii) a nine-dimensional model where the edge state is a periodic orbit and turbulence is represented by a chaotic saddle \cite{moehlis2004}, and eventually iv) well-resolved Navier-Stokes simulations of a periodic cell of plane Couette flow, understood as a $\mathcal{O}(10^5)$-dimensional dynamical system \cite{kawahara2005}. For all these models both LCS diagnostics display steep variations at the location of the edge manifold, which can be monitored using a single scalar quantity. Our findings indicate that having some prior knowledge of the system is helpful to properly identify the edge as an LCS, and that LDs are particularly sensitive to this. Approaching a model with unknown dynamical characteristics should be done in the following way: (1) the LCS diagnostic should be applied for the parameter region of interest (2) regions on both sides of a ridge in the FTLE or the boundaries between smooth or speckled regions of $B$ should be systematically explored. After testing different values of $\tau$, our exploration suggests that the edge manifold emerges already for $\tau=\mathcal{O}(1)$ but stands out more dramatically for longer time horizons, a feature worth a more quantitative study.

Furthermore the same toolbox allows, when it is unknown, to identify the edge state (the relative attractor on the edge manifold) using iterative bisection algorithms based on the LCS diagnostics, i.e.\ local measures in state space rather than global considerations as in previous works. The associated  algorithms have been tested on the Dauchot-Manneville model in two dimensions using both FTLEs and LDs: both outperformed the classical approaches, with a strong advantage for the LD-based methods in cases where the edge state is a fixed, point both in terms of scalability and computational cost. However, edge states are generally not fixed points.

As pointed out in \cite{Hadjighasem_2017} LDs, are not objective and their connection to LCS is unclear. This has lead to some controversy regarding the interpretation of the LD maps \cite{ruizherrera2015,ruizherrera2016,lopesino2017theoretical}. A hitherto unreported drawback of LDs, worth a more systematic characterisation, is that their ability to highlight the edge manifold depends on the dynamical nature of the invariant sets of the system. Consequently, their low computational cost comes with uncertainty about how to interpret the plots in an system where the nature of the attractors is unknown. The FTLEs represent a computationally more expensive tool. They have a solid mathematical background, are objective diagnostics \cite{Hadjighasem_2017} and with additional constraints can be precisely connected to hyperbolic LCSs \cite{Haller_2015}. Furthermore, they have a more direct interpretation in terms of material barriers.

The objectivity of the FTLEs makes them independent of the frame of reference, displaying the same values on the ridges and highlighting the edge as an LCS independently of the chosen frame. LDs are not objective and changing the frame of reference would imply a change in the values of the LD maps and even their nature. A relevant example is that of an edge state consisting of a travelling wave (TW) in physical space. Choosing a fixed frame of reference would yield a different result for the LD landscape than considering a frame moving with the velocity of the TW.

The present work paves the road for the design of effective local manifold tracking methods which do not rely on measurable global properties. The local edge tracking algorithms do not rely on a Boolean approach, but can be defined as optimisers of a scalar observable. We point out that the identification of local edge properties is not limited to bistable systems, and therefore as future work a revisit of manifold-tracking methods in different situations such as \cite{lebovitz2012boundary,zammert2019transition,beneitez2019edge} using this new framework is strongly encouraged.

\begin{acknowledgments}

Financial support by the Swedish Research Council (VR) grant no. 2016-03541 is gratefully acknowledged.
The authors thank H. Babaee for sharing the codes  from Ref. \cite{babaee2017} and for useful discussions. The late B. Eckhardt as well as P. Negi are acknowledged for useful discussions. Computing time provided by the Swedish National Infrastructure for Computing (SNIC) is gratefully acknowledged. The open source projects
\href{https://julialang.org/}{Julia},
\href{http://www.python.org}{Python},
\href{https://matplotlib.org/}{Matplotlib}
and
\href{https://www.paraview.org/}{ParaView}
have been used for this work.

\end{acknowledgments}


\bibliography{main.bib}

\end{document}